\documentclass[useAMS,usenatbib]{mnras}

\usepackage[T1]{fontenc}
\usepackage{ae,aecompl}
\usepackage{fnpct}

\usepackage{pdflscape}
\usepackage{graphicx}	
\usepackage{amsmath}	
\usepackage{amssymb}	
\usepackage{color}

\newcommand{\Msun}{M_{\sun}}
\newcommand{\kms}{km~s$^{-1}$}
\newcommand{\bmv}{\ensuremath{B\! - \!V}}

\newcommand{\mbh}{\ensuremath{M_{\rm BH}}}
\newcommand{\mbulge}{\ensuremath{M_{\rm bulge}}}
\newcommand{\mlk}{\ensuremath{\Upsilon_{K}}}
\newcommand{\mlt}{\ensuremath{\Upsilon_{\rm tot}}}
\newcommand{\mlkt}{\ensuremath{\Upsilon_{K,{\rm tot}}}}
\newcommand{\mlb}{\ensuremath{\Upsilon_{\rm bulge}}}

\newcommand{\mld}{\ensuremath{\Upsilon_{\rm disc}}}
\newcommand{\mlkd}{\ensuremath{\Upsilon_{K,{\rm disc}}}}
\newcommand{\msunlsun}{\ensuremath{M_{\odot}/L_{\odot}}}
\newcommand{\chisquare}{\ensuremath{\chi^{2}}}
\newcommand{\msigmarel}{\ensuremath{M_{\rm BH}}--\ensuremath{\sigma}}
\newcommand{\mbulgerel}{\ensuremath{M_{\rm BH}}--\ensuremath{M_{\rm bulge}}}
\newcommand{\dsoi}{\ensuremath{d_{\rm SoI}}}

\newcommand{\Imfit}{\textsc{Imfit}}

\title[Measuring SMBHs in Disc Galaxies: The Effects of Haloes]{NGC~307 and the Effects of Dark-Matter Haloes
on Measuring Supermassive Black Holes in Disc Galaxies}

\author[P. Erwin et al.]{Peter Erwin$^{1,2}$, Jens Thomas$^{1,2}$, 
Roberto P. Saglia$^{1,2}$, Maximilian Fabricius$^{1,2}$, 
\newauthor
Stephanie P. Rusli$^{1,2}$, Stella Seitz$^{1,2}$, and Ralf Bender$^{1,2}$  \\
$^{1}$Max-Planck-Insitut f\"{u}r extraterrestrische Physik, Giessenbachstrasse, 85748 Garching, Germany \\
$^{2}$Universit\"{a}ts-Sternwarte M\"{u}nchen, Scheinerstrasse 1, D-81679 M\"{u}nchen, Germany }

\begin{document}

\maketitle

\label{firstpage}

\begin{abstract} 

We present stellar-dynamical measurements of the central supermassive
black hole (SMBH) in the S0 galaxy NGC~307, using adaptive-optics IFU
data from VLT-SINFONI.  We investigate the effects of including
dark-matter haloes as well as multiple stellar components with different
mass-to-light ($M/L$) ratios in the dynamical modeling.  Models with no
halo and a single stellar component yield a relatively poor fit with a
low value for the SMBH mass ($7.0 \pm 1.0 \times 10^{7} \Msun$) and a
high stellar $M/L$ ratio ($\mlk = 1.3 \pm 0.1$). Adding a halo produces
a much better fit, with a significantly larger SMBH mass ($2.0 \pm 0.5
\times 10^{8} \Msun$) and a lower $M/L$ ratio ($\mlk = 1.1 \pm 0.1$). A
model with no halo but with separate bulge and disc components produces
a similarly good fit, with a slightly larger SMBH mass ($3.0 \pm 0.5
\times 10^{8} \Msun$) and an identical $M/L$ ratio for the bulge
component, though the disc $M/L$ ratio is biased high ($\mlkd = 1.9 \pm
0.1$). Adding a halo to the two-stellar-component model results in a
much more plausible disc $M/L$ ratio of $1.0 \pm 0.1$, but has only a
modest effect on the SMBH mass ($2.2 \pm 0.6 \times 10^{8} \Msun$) and
leaves the bulge $M/L$ ratio unchanged. This suggests that measuring
SMBH masses in disc galaxies using just a single stellar component and
no halo has the same drawbacks as it does for elliptical galaxies, but
also that reasonably accurate SMBH masses and bulge $M/L$ ratios can be
recovered (without the added computational expense of modeling haloes)
by using separate bulge and disc components.

\end{abstract}

\begin{keywords}
galaxies: structure -- galaxies: elliptical and lenticular, cD -- 
galaxies: bulges -- galaxies: individual: NGC~307 -- galaxies: evolution.
\end{keywords}

\section{Introduction} 

The most commonly used technique for measuring the masses of
supermassive black holes (SMBH) in galaxy centres is Schwarzschild
modeling; fully two-thirds of the SMBH masses in the recent compilations of
\citet{kormendy13} and \citet{saglia16} were determined this way.
Schwarzschild modeling entails the construction of gravitational
potentials based on the combination of a central SMBH and one or more
extended stellar components (which are typically based on deprojecting a
2D surface-brightness model of the galaxy in question), with the SMBH
mass and stellar mass-to-light ($M/L$) ratio as variables. A library of
stellar orbits is built up by integrating test particles within a given
potential defined by particular values of SMBH mass and stellar $M/L$
ratio; these orbits are then individually weighted so as to reproduce
the observed light distribution and stellar kinematics of the galaxy.
The SMBH mass and stellar $M/L$ ratio are varied until the best match
with the data is achieved.

Schwarzschild modeling has several advantages over methods based on
modeling gas kinematics (the other major approach for measuring SMBH
masses): it can be used in any galaxy bright enough for stellar
kinematics to be measured, does not require the presence of gas, and
does not require simplifying assumptions about the underlying kinematics
(e.g., that all orbits are circular and coplanar).

Up until recently, the standard approach for Schwarzschild
modeling of SMBH masses has been to treat galaxies as having just two
components: a central SMBH and a stellar component with a single $M/L$
ratio. This is problematic for several reasons, the principal ones being
that galaxies -- especially disc galaxies -- do not always have uniform
$M/L$ ratios, and that galaxies have dark matter as well as stars. 

Disc galaxies are widely recognized as having spatially varying stellar
$M/L$ ratios, something at least partly due to different stellar
populations in different subcomponents. \citet{davies06} introduced the
idea of using two stellar components with distinct, independent $M/L$
ratios in order to model the combination of an actively star-forming
nuclear star cluster within an older bulge in the spiral galaxy
NGC~3227. \citet{nowak10} modeled the central bulges and main discs as
separate stellar components for two spiral galaxies (NGC~3368 and
NGC~3489); this was also done by \citet{rusli11} for the S0 galaxy
NGC~1332. The modeling of separate $M/L$ ratios for bulges is also
useful for investigating bulge-SMBH correlations, especially if one
wants to determine bulge masses \textit{dynamically}
\citep[e.g.][]{haring04,saglia16}.

Elliptical galaxies are in principle simpler to model than disc
galaxies,  because we can treat ellipticals as having a single stellar
component (i.e., they can be approximated as pure ``bulge'' with no
disc). However, they are known -- like all galaxies -- to possess haloes
of dark matter. Recent work has focused on the question of whether the
practice of ignoring these haloes in dynamical modeling might bias the
resulting SMBH masses and stellar $M/L$ ratios. The key issue is whether
the modeling process assigns extra mass to the stellar component in
order to account for the (missing) effect of the halo. An increased
stellar $M/L$ ratio can then result in a lower SMBH mass, because the
stars at small radii will contribute more to the central potential than
they would if the $M/L$ ratio were lower; this removes the need for a
more massive SMBH.

\citet{gebhardt09} found that including a DM halo in their models for
M87 resulted in a stellar $M/L$ ratio about half as large -- and a SMBH
mass about twice as large -- as when their models included only a SMBH
and the stellar component.  Subsequent studies examining the inclusion
of DM haloes in elliptical-galaxy models have yielded somewhat
conflicting results, with some reporting effects similar to those found
by \citet{gebhardt09} -- e.g., \citet{mcconnell11a} -- and some
reporting no differences between models with and without DM haloes --
e.g., \citet{shen10a,jardel11}.\footnote{The \citet{jardel11} study is
of the bulge-dominated Sa galaxy NGC~4594, not an elliptical.} Studies
of larger samples by \citet{schulze11} and \citet{rusli13a} have
indicated that DM haloes can be safely ignored in the modeling
\textit{only} if high-spatial-resolution kinematics are available for
the centre of the galaxy. Ideally, this means kinematic observations
obtained with a point-spread-function whose FWHM is at least 5--10 times
smaller than the diameter of the SMBH's sphere of influence
\citep{rusli13a}.

What is not clear at this point is whether ignoring the existence of
dark matter haloes in dynamical models of \textit{disc} galaxies has any
significant effect on either derived SMBH masses or bulge $M/L$
ratios. In this paper, we investigate this question by measuring the central
SMBH mass and stellar $M/L$ ratios for the S0 galaxy NGC~307 using a
four different models: first, a simple SMBH + single-stellar-component
model; second, a model with a SMBH and \textit{two} stellar components
(bulge and disc) with separate $M/L$ ratios. We then add dark-matter haloes
to both the single- and two-stellar-component models.

Unless otherwise specified, we adopt a cosmology where $\Omega_{m} = 0.7$,
$\Omega_{\Lambda} = 0.3$, and $H_{0} = 75$ \kms{} kpc$^{-1}$.

\begin{figure}
\begin{center}
\includegraphics[scale=1.05]{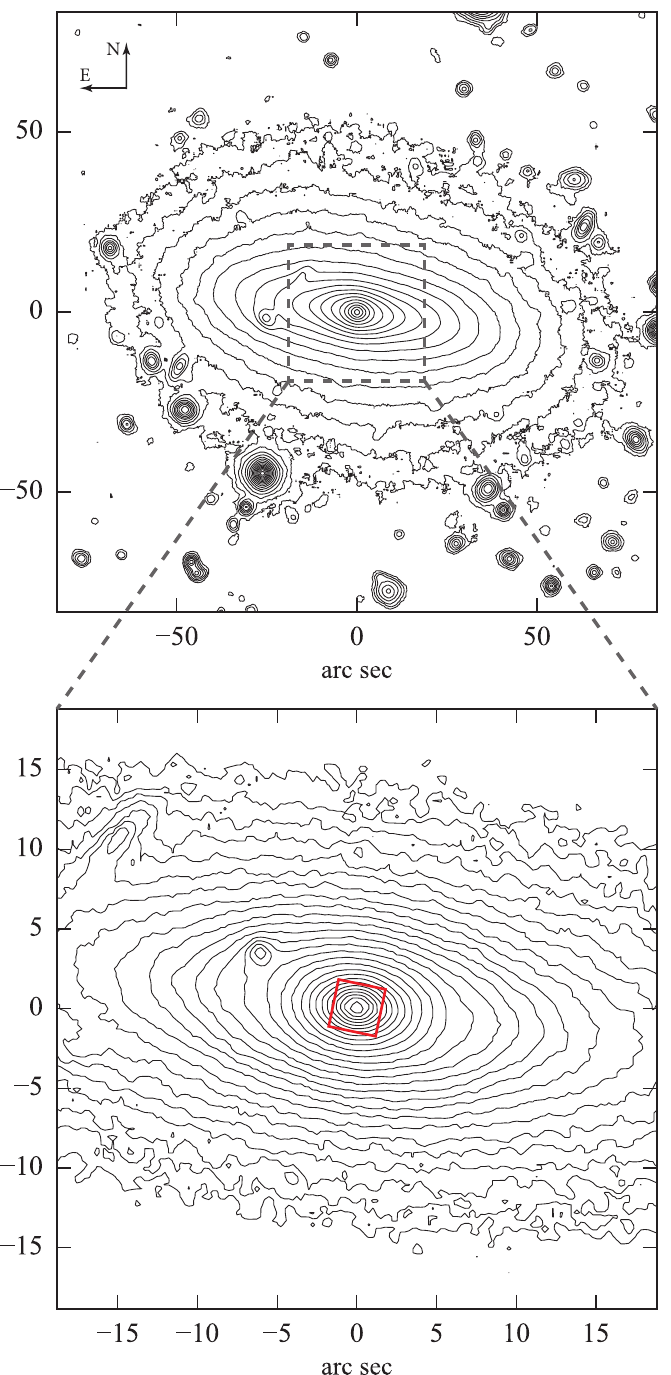}
\end{center}

\caption{\textbf{Top:} Logarithmically scaled isophotes for the $R$-band
WFI image of NGC~307 (smoothed with a 9-pixel-wide median filter). An
elliptical disc embedded within a rounder and slightly skewed stellar halo
can be seen.
\textbf{Bottom:} Close-up of VLT-FORS1 $R$-band image (smoothed with a
3-pixel-wide median filter), showing the rounder bulge region within
the disc. The small red square indicates the
approximate field of view and orientation of our SINFONI observation.
\label{fig:n307-isophotes}}

\end{figure}

\section{NGC 307}\label{sec:n307}

NGC 307 is a poorly-studied early-type galaxy, classified as S$0^{0}$ by
\citet{rc3}. Although it lies only $\sim 0.5\degr$ from the centre of
the cluster Abell~119, its much smaller redshift (0.0134 versus 0.044
for the cluster) means there is no physical association. In the group
catalog of \citet{garcia93}, it is the second-brightest\footnote{Based
on tabulated values in NED.} member of a small, five-galaxy group (LGG
13, brightest member = NGC~271). We adopt a distance of 52.8 Mpc, based
on the (Virgocentric-infall-corrected) redshift of 3959 \kms{} from
HyperLEDA. \citet{tonry81} reported a central velocity dispersion of
$325 \pm 15$ \kms{}, but more modern measurements indicate significantly
lower values: $\sigma_e = 239$ \kms{} has been reported by
\citet{vandenbosch15}, and \citet{saglia16} estimated $\sigma_{e} = 205$
\kms, based on the kinematic and imaging data presented in this
paper.\footnote{\citet{saglia16} used a curve-of-growth analysis of the
VLT-FORS1 image to derive a whole-galaxy $r_{e} = 4.76\arcsec$; the
light-weighted dispersion within this radius was determined as described
in Appendix~A of that paper, using the VLT-FORS1 long-slit data.} Using
the HyperLeda corrected $\bmv$ colour (0.84) and the colour-based $M/L$
ratios of \citet{bell03} with either the HyperLeda $B_{\rm tc}$
magnitude (13.52) or the 2MASS total $H$ magnitude
(9.865),\footnote{Corrected for Galactic extinction using data from
\citet{schlafly11}, as tabulated in NED.} we find estimated stellar
masses of either $5.5 \times 10^{10} \Msun$ or $6.5 \times 10^{10}
\Msun$, quite close to recent estimates of the Milky Way's stellar mass
\citep[e.g.,][]{mcmillan11,licquia15,mcmillan17}.

\begin{figure}
\begin{center}
\hspace*{-15mm}
\includegraphics[scale=0.59]{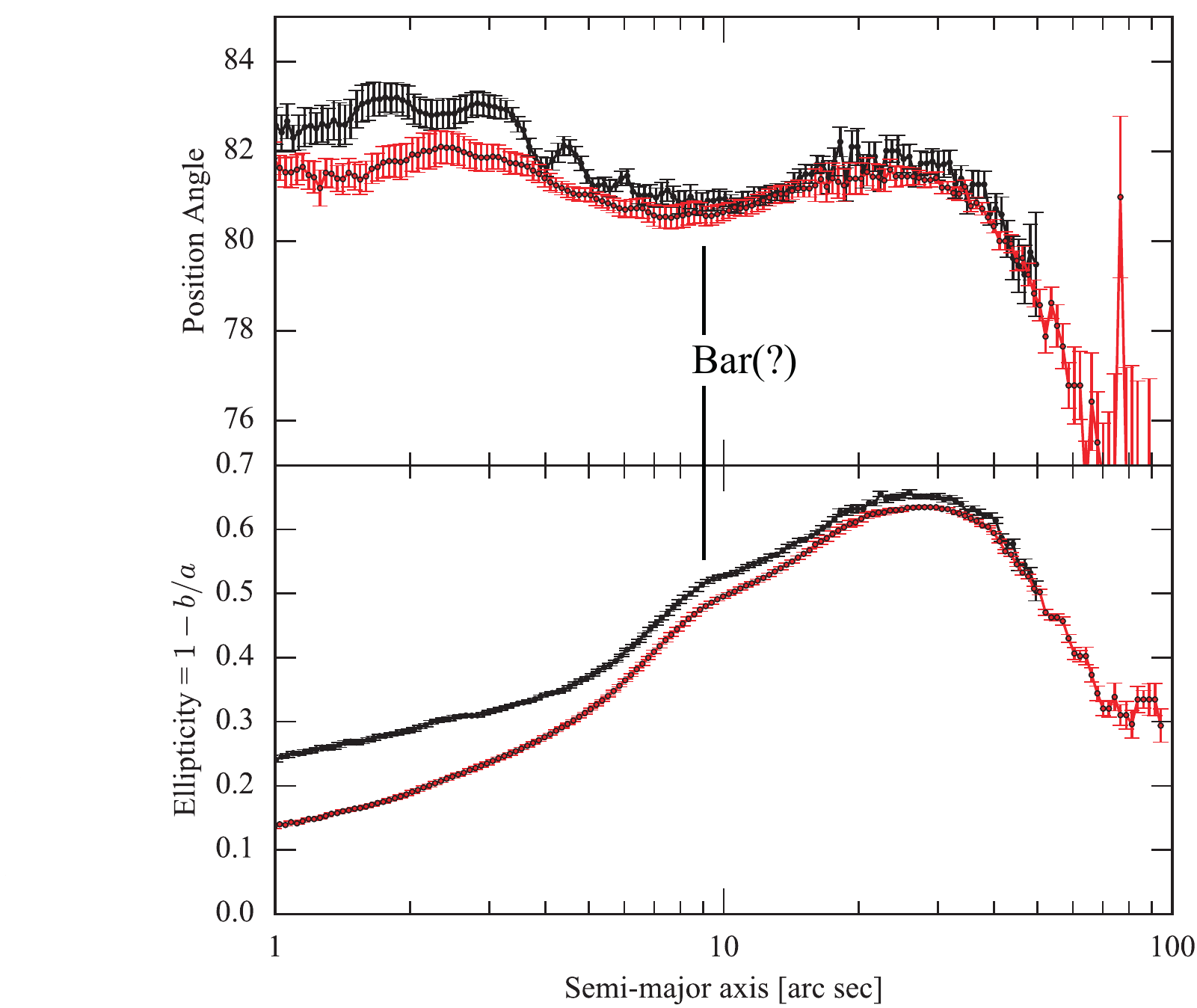}
\end{center}

\caption{Position angles and ellipticities from ellipse fits to the WFI
(red) and VLT-FORS1 (black) $R$-band images of NGC~307. The label
indicates the position-angle twist and local shoulder in the ellipticity
profile corresponding to a possible bar or lens. \label{fig:n307-efits}}

\end{figure}

Figure~\ref{fig:n307-isophotes} shows log-scaled $R$-band isophotes of
NGC~307 using an image from the Wide Field Imager (WFI) on the ESO 2.2m
telescope and a higher-resolution image from the FORS1 imager-spectrograph
on the VLT; ellipse fits to
both images are shown in Figure~\ref{fig:n307-efits} (see
Section~\ref{sec:imaging-data} for more about the images). These fits
show a fairly broad peak in isophotal ellipticity of $\epsilon \approx
0.65$ extending from semi-major axis $a \sim 20\arcsec$ to $a \sim 40\arcsec$,
with the isophotes becoming significantly rounder (as low as $\sim
0.30$) further out. This suggests that we may be seeing a disc embedded
within a rounder, luminous halo (we will show in
Section~\ref{sec:2dfits} that the latter is unlikely to be just an extension
of the central bulge). In addition, unsharp masks suggest the existence
of a weak bar or lens within the disc, with semi-major axis $\sim
10\arcsec$; this matches the shoulder in ellipticity seen in the ellipse
fits and the corresponding slight twist in the position angle to a local
minimum of $\sim 81\degr$ at $\approx 9$--10\arcsec.

\section{Observations} 
\label{sec:obs}

\subsection{Spectroscopy: SINFONI IFU Data} 
\label{sec:sinfoni-data}

Our primary set of spectroscopic data comes from observations made at
the VLT with SINFONI in November of 2008. SINFONI combines the near-IR
integral field spectrograph SPIFFI and the adaptive-optical module MACAO
\citep{eisenhauer03,bonnet04}, using an image-slicer to subdivide the
field of view into 32 slitlets, which are subsequently rearranged into a
composite pseudo-long-slit image that is passed into the main
spectrograph.  After dispersion by the grating, the resulting composite
spectrum is imaged onto a $2048 \times 2048$ Hawaii 2RG detector.

The pre-optics of SINFONI allow the user to select one of three
different spatial resolution modes: 25, 100, or 250 mas, corresponding
to fields of view of $0.8\arcsec \times 0.8\arcsec$, $3\arcsec \times
3\arcsec$, or $8\arcsec \times 8\arcsec$.  For NGC~307, we used only the
middle (100mas) scale, along with the $K$-band grating, since our
primary target was the CO absorption bandheads at $2.3$~\micron. A
single exposure, when assembled into a datacube, yields rectangular
spatial elements with sizes of $50 \times 100$ mas for the 100 mas mode;
when multiple exposures with appropriate dithering are combined, the
resulting datacubes have a spatial pixel scale of 50 mas pixel$^{-1}$.
The resulting $K$-band velocity resolution is $\sigma = 53$ \kms.

Since NGC~307 is much larger than the SINFONI field of view, we
observed it using a sequence of multiple ten-minute exposures organized
into an object-sky-object pattern; the sky exposures were made with
an offset of 80\arcsec{} along the galaxy minor axis to avoid
contamination by galaxy light. Individual ten-minute exposures were
dithered using offsets of a few (spatial) pixels, to reduce the effects
of bad pixels in the detector and to allow construction of a final data
cube with full spatial resolution. The complete set of observations
included 40 minutes of on-target time on each of two night -- 2008
November 25 and 26 -- for a total of 80 minutes integration time.
However, we found the observations from the first night to be of
significantly higher quality in terms of AO performance and achieved
resolution; since they had sufficient S/N by themselves, we only used
that night's data. Observations of telluric-standard B stars, obtained
immediately after the galaxy observations and at similar air masses,
were used to remove atmospheric absorption (see below).

The centre of NGC~307 was not bright enough to serve as an AO guide
source by itself, so we used the PARSEC laser guide star (LGS)
system at the VLT \citep{bonaccini02,rabien04}. The LGS mode still
requires an extra-atmospheric reference source for ``tip-tilt''
correction of lower-order atmospheric distortions; we used the galaxy
nucleus for this.

Data reduction was performed using a custom-built pipeline combining the
official ESO SINFONI Pipeline \citep{modigliani07} with elements from
its predecessor, the SPIFFI Data Reduction Software
\citep[SPRED;][]{schreiber04,abuter06}. This combined pipeline included
the standard bias-correction, dark subtraction, distortion correction,
non-linearity correction, flat-fielding, wavelength calibration, and
datacube generation stages. Sky subtraction, which used the sky datacube
observed closest in time for each galaxy datacube, was augmented using
the IDL code of \citet{davies07a} to account for variations in night-sky
emission-line strengths between the times of the galaxy and sky
observations.

The galaxy datacubes were then corrected for telluric absorption using
the telluric-star datacubes. This involved extracting a single, summed
spectrum for the telluric star from its datacube and then dividing it by
a blackbody curve with a temperature appropriate for the spectral type
of the star (the Paschen $\gamma$ absorption line in the standard-star
spectrum was fit by hand using code written in IDL). The resulting
normalized spectrum was then used to correct the individual spectra in
the corresponding galaxy datacubes. Finally, the individual datacubes
were combined into a single datacube for the night, taking into account
the recorded dither positions in the headers. 

To estimate the resolution obtained by the LGS system, we used ``PSF
star'' observations obtained during or immediately after the  galaxy
observations, with exactly the same instrument setup and AO mode (i.e.,
LGS). To make the match as close as possible, the PSF stars were chosen
to have the same $R$-band magnitude and $B - R$ colour as the galaxy
nucleus (measured within a 3\arcsec-diameter aperture), so that the AO
system would respond in a similar fashion. Although it is always
possible that the PSF star measurements reflect different observing
conditions, \citet{hicks13} reported that measurements of PSF stars
taken after their (non-AO) VLT-SINFONI observations showed FWHM
agreement to within 0.02\arcsec{} for galaxies with bright AGN, where
the AGN itself could be used to independently measure the seeing. The
combined PSF-star datacube was flattened to produce a $K$-band image,
which was then fit with the sum of two Gaussians using \textsc{Imfit}
\citep{erwin15-imfit}. The inner Gaussian component (37\% of the total
light), which was mildly elliptical, had FWHM measured along its major
and minor axes of 0.20\arcsec{} and 0.16\arcsec{}, respectively, for a
mean resolution of 0.18\arcsec. The outer component was nearly circular,
with a FWHM of 0.48\arcsec. This PSF is consistent with
previously published SINFONI 100mas K-band values when using the laser
guide star; in fact, it is equal to the median value from our previously
published observations with the LGS in the same mode
\citep{nowak10,rusli13a,mazzalay13}.

\subsection{Spectroscopy: VLT/FORS1 and VIRUS-W Observations}

To obtain measurements of the stellar kinematics outside the central
$3\arcsec \times 3\arcsec$ field of view provided by our SINFONI
data, we made two sets of optical spectroscopic measurements: long-slit
observations along the galaxy major and minor axes with the FORS1
spectrograph in the VLT, and wide-field IFU observations with the VIRUS-W 
spectrograph on the McDonald 2.7m telescope.

\subsubsection{VLT/FORS1}


We obtained long-slit data along the galaxy major and minor axes with
the VLT-FORS1 spectrograph on 2008 October 23 (Programme ID 082.A-0270).
We made a total of four 2700s exposures with the slit oriented along the
galaxy major axis (PA = 78.1\degr) and two more exposures of the same
integration time with the slit along the minor axis (PA = 168.1\degr).
The instrument was used with the 1200g grism and a slit width of
1.6\arcsec{} width slit; the instrumental dispersion was $\sigma \approx
79$ \kms.

The reduction of the FORS1 spectra followed the standard steps of bias
subtraction, flat fielding, cosmic-ray rejection, and wavelength
calibration to a logarithmic scale using our customized MIDAS scripts
\citep{delorenzi08}. We subtracted the sky measured at the ends
of the slit and binned the resulting frame radially to obtain a set of
spectra with approximatly the same signal to noise ratios.  

The kinematic analysis of the spectra is discussed in in
Section~\ref{sec:optical-kinematics}, and our stellar-population
analysis is discussed in Section~\ref{sec:stellar-pop}.

\subsubsection{VIRUS-W}

VIRUS-W is an optical integral-field-unit spectrograph with a
$105\arcsec \times 55\arcsec$ field of view, based on the VIRUS IFU
design for HETDEX (Hoby-Eberly Telescope Dark Energy Experiment) and
adapted to achieve high spectral resolution for deriving stellar
kinematics \citep{fabricius12b}. It has 267 fibers with core diameters
of 3.14\arcsec{} on the sky, arranged in a rectangular array with a fill
factor of one-third. 

We observed NGC~307 with VIRUS-W mounted on the 2.7m Harlan J. Smith
telescope at the McDonald Observatory in Texas on 2010 December 6, as
part of commissioning/science-verification time for the instrument. The
galaxy was observed using a total of three dither positions (to account
for the 1/3 fill factor), each with 1200s exposure time. These were
bracketed and interleaved with sky offset exposures, also using 1200s
exposure times. The seeing varied in FWHM from 1.2\arcsec{} to 1.9\arcsec. VIRUS-W
has both low- and high-resolution spectral modes; although we observed
NGC~307 with both modes, we ended up using just the low-resolution mode
data. Since the low-resolution mode has $\sigma_{\rm instr} = 39$ \kms{}
($R = 3300$, with a spectral coverage of 4320--6042 \AA), it provided
more than sufficient spectral resolution for NGC~307.

Data reduction used a custom pipeline based on the Cure pipeline for
for HETDEX; see \citet{fabricius14} for details.  The result is a datacube
with $1.6 \times 1.6\arcsec$ spaxels.

In order to generate high-$S/N$ spectra for kinematic extraction, we
combined spectra from individual spaxels using the Voronoi binning
scheme of \citet{cappellari03}, ending up with a median $S/N$ per bin of
29. The kinematic analysis of the binned spectra is discussed in in
Section~\ref{sec:optical-kinematics}.

\subsection{Imaging Data}\label{sec:imaging-data} 

The available imaging data for NGC~307 consist of a large-scale $R$-band
image from a 300s exposure at the ESO-MPI 2.2m Wide Field Imager on 2010
July 15 (Programme ID 084.A-9002), with seeing FWHM $= 1.62\arcsec$; a
10s-exposure VLT-FORS1 image with smaller field of view (also $R$-band,
with FWHM = 1.00\arcsec) made during our spectroscopic observations with
FORS1 (above); and our VLT-SINFONI combined datacube, collapsed along
the wavelength axis to form a $\sim 3\arcsec \times 3\arcsec$ $K$-band
image. 

These images are, to a degree, complementary: the WFI image is wide
enough and deep enough to allow determination of the outer stellar halo
and main disc, but has relatively poor resolution; the FORS1 image
provides better resolution for the bar/lens and the disc-bulge
transition region, but is not as good for characterizing the halo due to
its smaller field of view, lower S/N, and the fact that the outer part
of the galaxy falls on an inter-chip gap; and the SINFONI image
has the best resolution for the inner region of the bulge. Consequently,
we construct our final photometric models using a combination of all
three images. 

Since the innermost data are $K$-band, we calibrated all three images to
$K$-band by a multi-step process similar to that used by
\citet{nowak10}; the resulting calibration is ultimately based on the
publicly available 2MASS $K$-band image of the galaxy. First, we
calibrated the FORS1 image by convolving it to the resolution of the
2MASS image and performing aperture photometry on both images. We then
calibrated the SINFONI $K$-band image to the FORS1 image by iteratively
matching surface-brightness profiles from ellipse fits to both images in
the region $a = 0.6$--1.42\arcsec, including a sky-background term for
the SINFONI data.\footnote{This is because of variations in the sky
background between the times of the galaxy and sky observations with
SINFONI, which cannot be completely removed by the data reduction
process.} Finally, the WFI image was calibrated to match the
$K$-band-calibrated FORS1 image using a similar ellipse-fit
profile-matching technique for the region $a = 15$--45\arcsec.

\section{Stellar Kinematics}

\subsection{SINFONI Kinematics}\label{sec:sinfoni-kinematics}

For our SINFONI data, we extracted full, non-parametric line-of-sight
velocity distributions (LOSVDs) from the spectra, using a total of 21
bins in velocity space. We used a maximum penalized likelihood (MPL)
method originally introduced by \citet{gebhardt00b} and a set of stellar
template spectra of K and M giants derived from earlier SINFONI
observations with the same instrumental setup \citep[see,
e.g.][]{nowak07,nowak08,nowak10}.\footnote{The extreme width of the CO
bandheads makes the FCQ method we use for our optical spectra
(Section~\ref{sec:optical-kinematics}) unusable.}

\begin{figure}
\begin{center}
\includegraphics[scale=0.95]{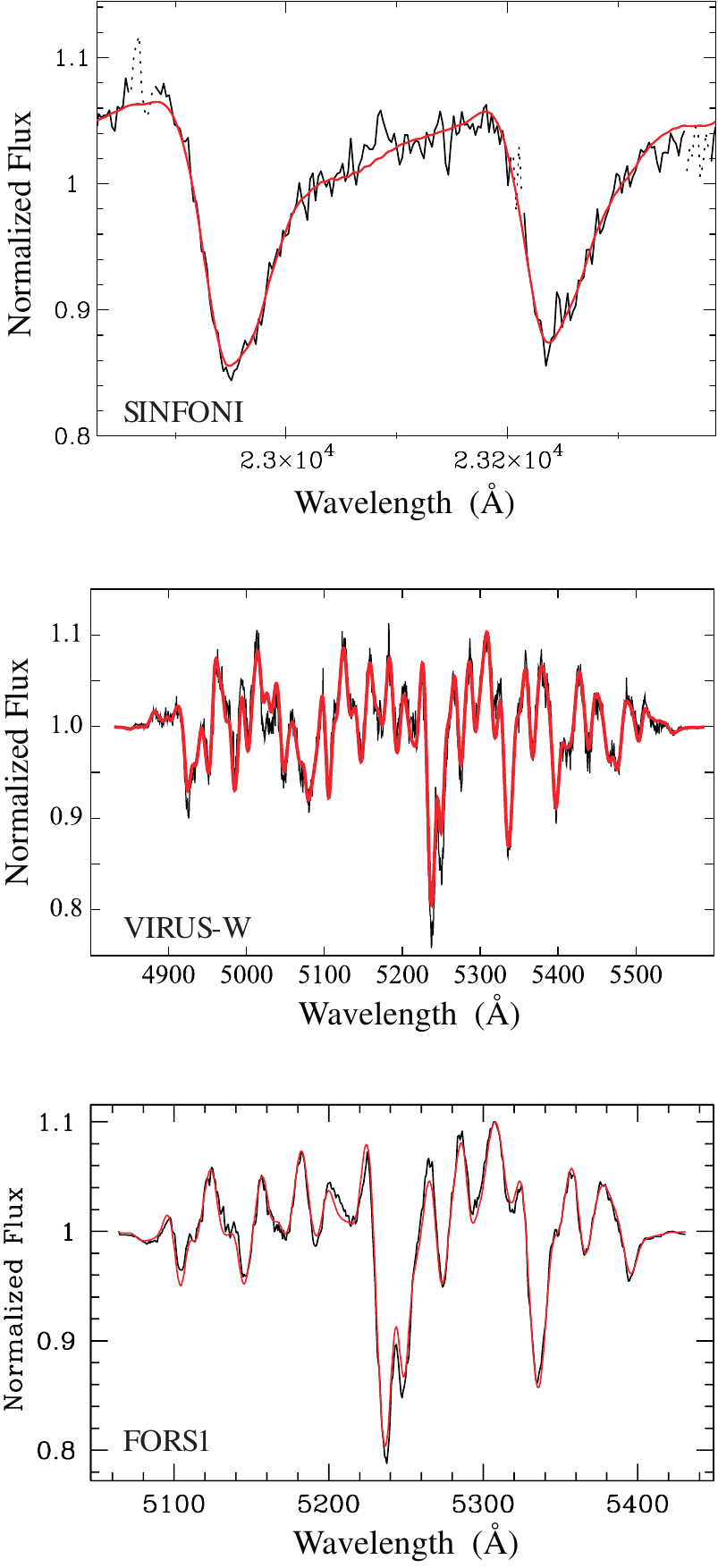}
\end{center}

\caption{Examples of kinematic fits to our spectroscopy. Top: best-fit,
LOSVD-convolved model spectrum (red) and binned VLT-SINFONI data from
one of the central bins (black line); dashed lines indicate regions of the spectrum
not used in the fit. The observed spectrum has been normalized by division
by a smooth continuum fit. Middle: Best-fit model spectrum (red) and observed
optical spectrum (black) from the central bin of the VIRUS-W observations; the
observed spectrum has been normalized by subtracting a smooth continuum fit.
Bottom: Same, but now for the FORS1 major-axis spectrum.
\label{fig:spectra-fits}}

\end{figure}

We focused on the spectral region containing the first two CO bandheads
$^{12}$CO(2--0) and $^{12}$CO(3--1), which corresponds to a rest-frame
spectral range of 2.279--2.340 \micron. In order to minimize template
mismatch, we limited our set of template stars to those with
equivalent widths for the first CO bandhead which were similar to the 
equivalent width of the galaxy spectra \citep{silge03}. A trial LOSVD was
convolved with a linear combination of template spectra, and the
resulting model spectrum was compared with the data. The LOSVD and the
weights for the template spectra were adjusted by minimizing a penalized
\chisquare{} function:
\begin{equation}
\chi^{2}_{\rm P} \; = \; \chi^{2} \, + \, \alpha \mathcal{P},
\end{equation}
where $\mathcal{P}$ is the penalty function (the integral of
squared second derivative of the LOSVD) and $\alpha$ is a smoothing
parameter. The appropriate value of $\alpha$ depends on the S/N of
the data and the velocity dispersion of the galaxy; our choice was
based on extensive simulations involving MPL fitting of template stellar spectra
convolved with different LOSVDs; see \citet{nowak08} for more details.
An example of one of our fits is shown in the upper panel of Figure~\ref{fig:spectra-fits}.

\begin{figure*}
\begin{center}
\includegraphics[scale=1.05]{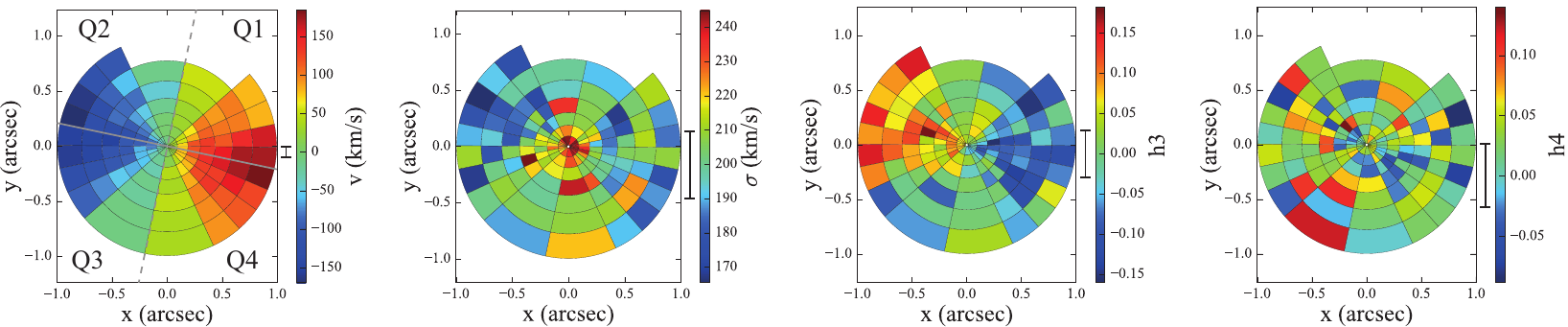}
\end{center}

\caption{Stellar kinematics (from left to right: velocity, velocity
dispersion, $h_{3}$, and $h_{4}$) from our VLT-SINFONI observations of
NGC~307, using our radial and angular binning scheme. Note that our
dynamical modeling uses the full LOSVD from each bin, not the
Gauss-Hermite moments we show in this figure. Maps have been rotated
so that north is up and east is to the left. Solid and dashed grey
lines in the first panel indicate galaxy major and minor axes, respectively;
``Q1'' through ``Q4'' labels indicate Quadrants 1 through 4, as used
in our dynamical modeling.
Error bars next to the colour bars indicate median errors from Monte
Carlo simulations.
\label{fig:sinfoni-kinematics}}

\end{figure*}

To increase the S/N of the spectra, we binned individual spaxels into
angular and radial bins using luminosity-weighted averaging. This
involved dividing the galaxy into four quadrants; the boundaries of the
quadrants were set by the major and minor axes of the galaxy. Each
quadrant was subdivided into five angular bins and seven radial bins. (See
Figure~\ref{fig:sinfoni-kinematics} for the binning, and the first panel
for definitions of the quadrants.)

Uncertainties for the best-fitting LOSVDs were determined by a Monte Carlo
technique, where for each spectrum we created 100 realizations of the
best-fitting combined template spectrum, convolved with the best-fitting
LOSVD, and then added Gaussian noise based on the measured RMS
deviations of the original fit. Each such spectrum was then fit using
the same MPL approach, with the final uncertainties based on the
distribution of fitted LOSVDs from the Monte Carlo realizations.

For presentation purposes, we parameterized the LOSVDs using the Gauss-Hermite
moments \citep{gerhard93,van-der-marel93} velocity $v$, velocity
dispersion $\sigma$, $h_{3}$, and $h_{4}$. Maps of these four moments
are shown in Figure~\ref{fig:sinfoni-kinematics}. Significant rotation can
be seen in the velocity field, with an accompanying anti-correlation in the
$h_{3}$ values. A somewhat noisy trend of increasing velocity dispersion towards the
galaxy centre can also be seen. No trends are visible in the $h_{4}$ map.

\subsection{Optical Kinematics}\label{sec:optical-kinematics}

Stellar kinematics for both the VLT-FORS1 long-slit spectra and the
Voronoi-binned VIRUS-W spectra were derived using the Fourier
Correlation Quotient (FCQ) method \citep{bender90,bender94}, which
models the LOSVD using a Gauss-Hermite decomposition, producing stellar
velocity $V$ and velocity dispersion $\sigma$ values, along with $h_{3}$
(skew) and $h_{4}$ (kurtosis) deviations from Gaussianity. The FORS1
kinematics were measured as done in \citet{saglia10}, chosing the
best-fitting template from the simple stellar population model spectra
of \citet{vazdekis99}. For the VIRUS-W data, we used a single K2 III
template star (HR 2600) spectrum previously observed with VIRUS-W, using
a rest-frame spectral range of 4537--5442 \AA{} and removing the
continuum using an eight-order polynomial. Error estimates for the $V$,
$\sigma$, $h_{3}$, and $h_{4}$ measurements in both cases were obtained
using a Monte Carlo approach \citep{mehlert00}. Examples of individual
fits are shown in the middle and lower panels of Figure~\ref{fig:spectra-fits}.

\begin{figure*}
\begin{center}
\includegraphics[scale=0.93]{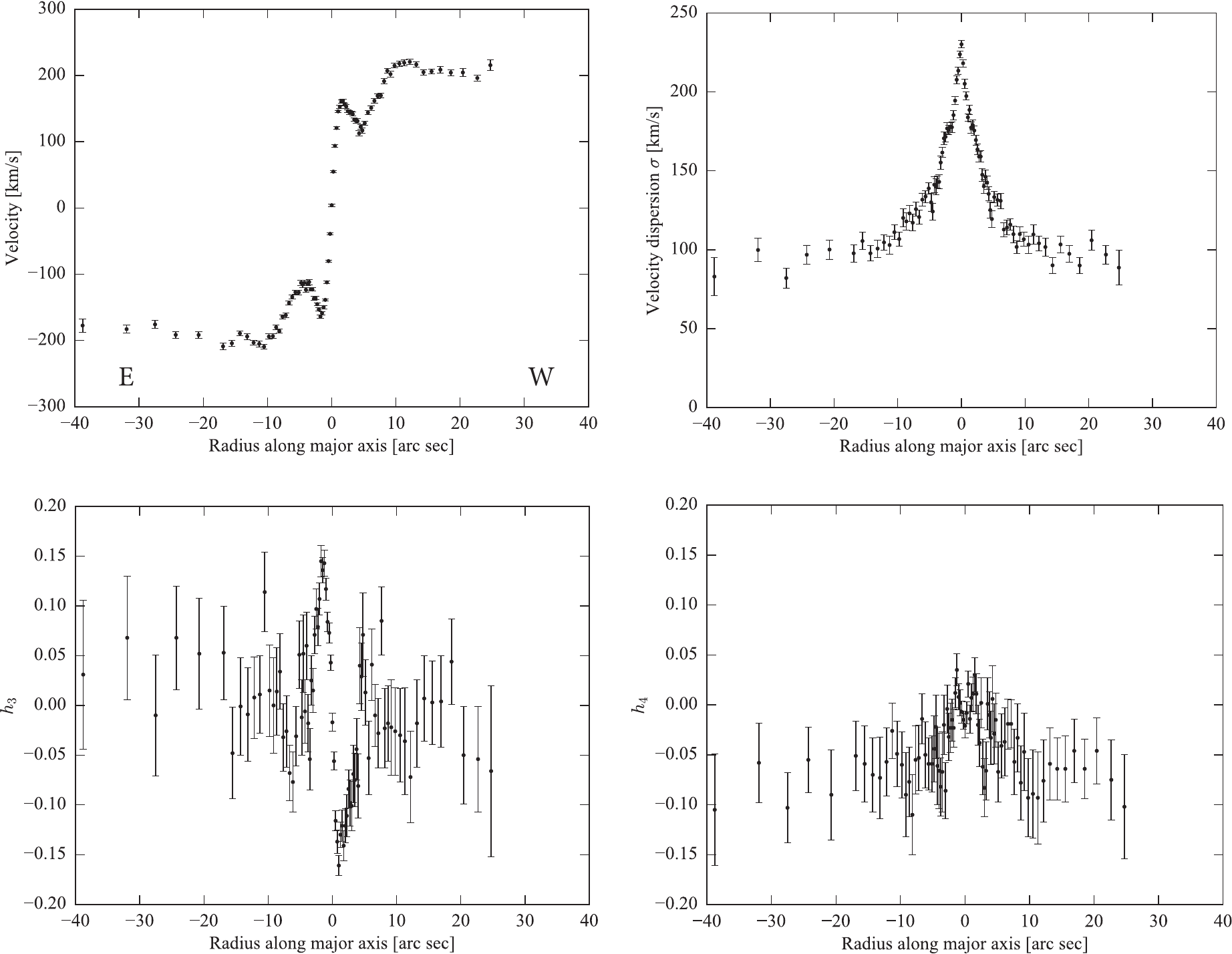}
\end{center}

\caption{Major-axis (PA $= 78.1\degr$) stellar kinematics from our VLT-FORS1 observations of NGC~307.
\label{fig:n307-fors1kin-major}}

\end{figure*}

\begin{figure*}
\begin{center}
\includegraphics[scale=0.93]{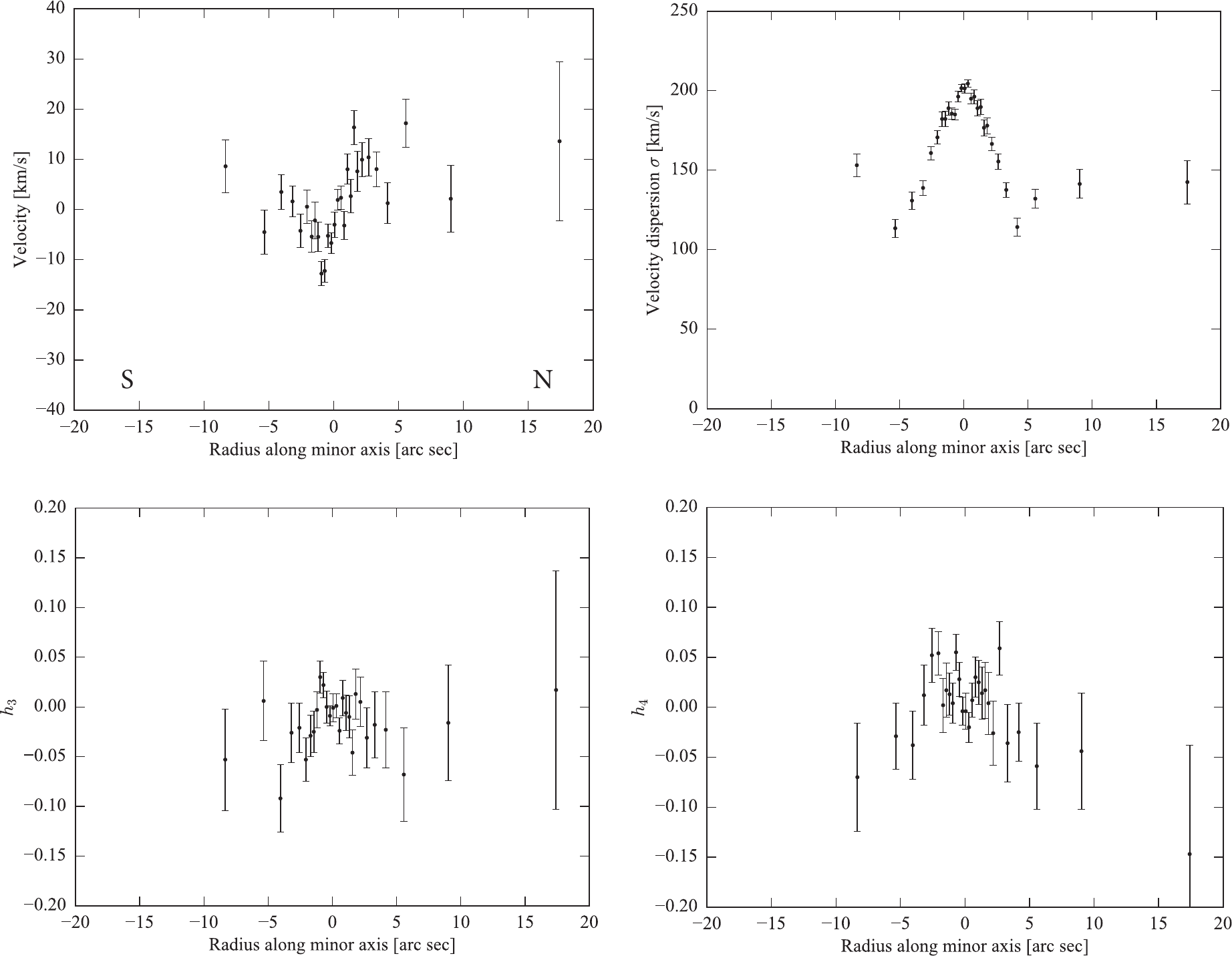}
\end{center}

\caption{Minor-axis (PA $= 168.1\degr$) stellar kinematics from our VLT-FORS1 observations of NGC~307.
\label{fig:n307-fors1kin-minor}}

\end{figure*}

\begin{figure*}
\begin{center}
\includegraphics[scale=0.6]{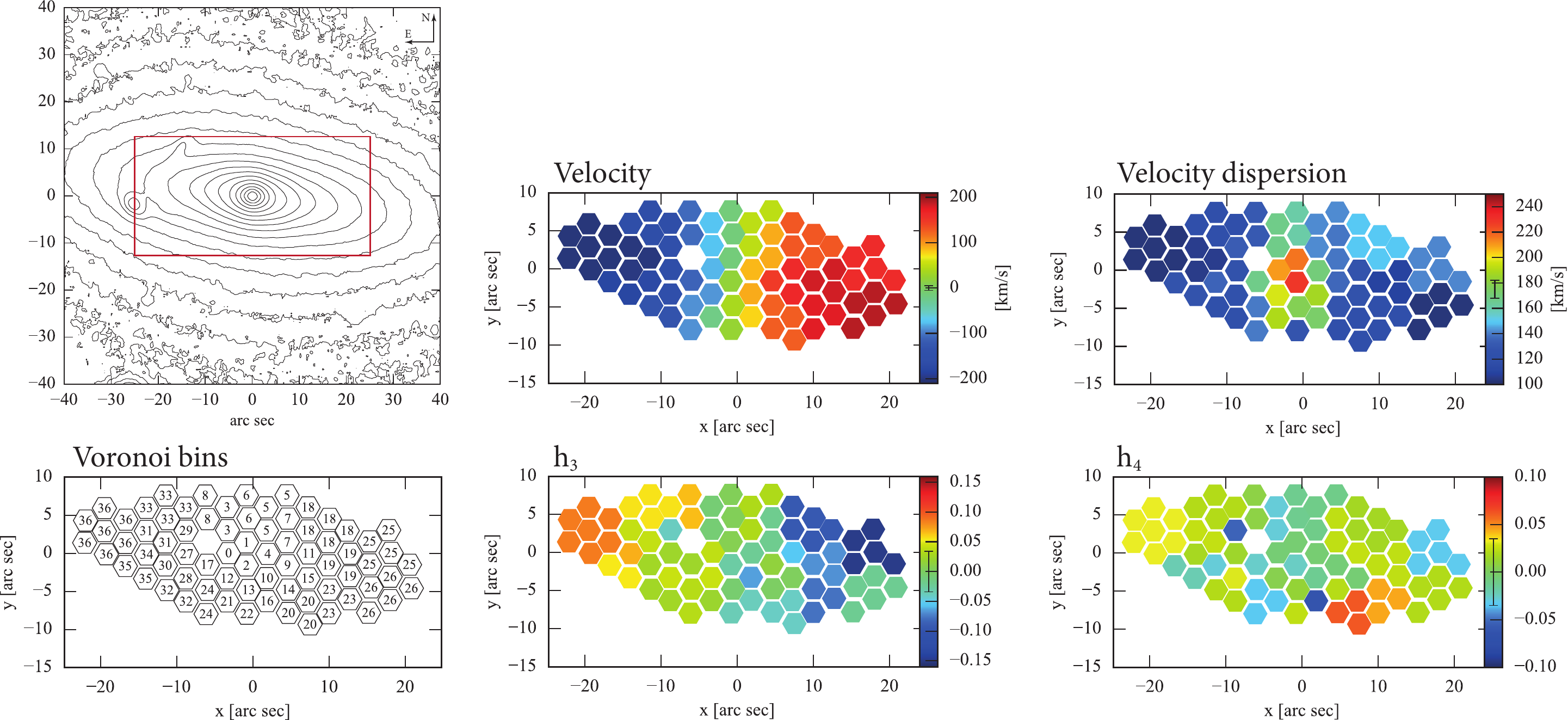}
\end{center}

\caption{Stellar kinematics from our VIRUS-W observations of NGC~307.
\textbf{Upper left:} $R$-band contours for NGC~307 from WFI image
(median-smoothed with width = 5 pixels); the red box corresponds to the sizes 
of the other panels. \textbf{Lower left:} Map of fiber positions; numbers
indicate which Voronoi bins individual fibers belong to. \textbf{Middle
and right:} Stellar kinematic maps ($V$, $\sigma$, $h_{3}$, $h_{4}$).
Error bars inside the colour bars indicate median errors from Monte
Carlo simulations. \label{fig:n307-virus-kinematics}}

\end{figure*}

\begin{figure*}
\begin{center}
\hspace*{-4mm}
\includegraphics[scale=0.8]{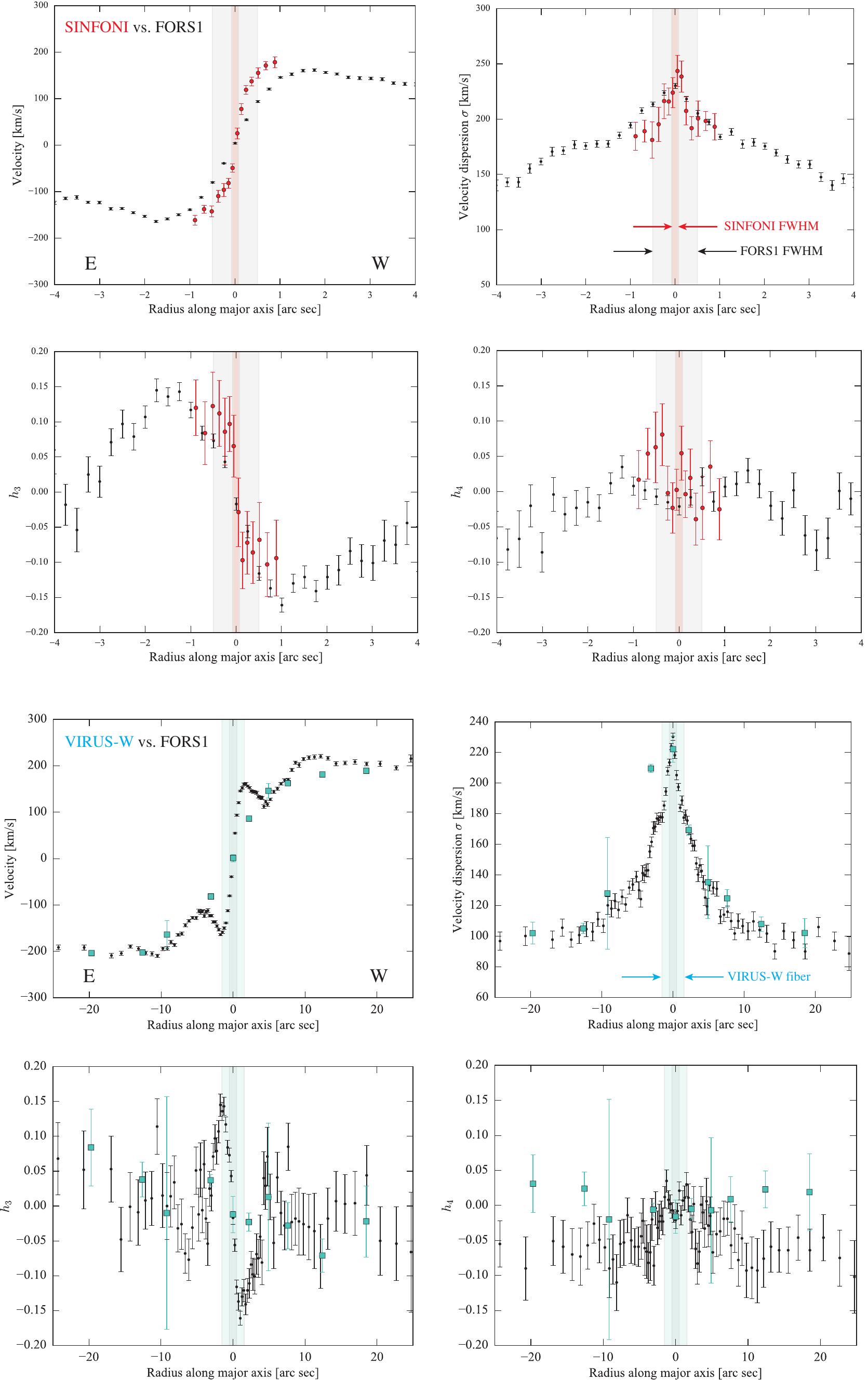}
\end{center}

\caption{Comparison of major-axis stellar kinematics from our SINFONI (medium-sized red circles), VLT-FORS1 (small black circles), and VIRUS-W (cyan squares) observations of NGC~307. The FWHM or fiber sizes of the observations are indicated by
the vertical shaded regions.
\label{fig:n307-compare-kinematics}}

\end{figure*}

Figures~\ref{fig:n307-fors1kin-major} and \ref{fig:n307-fors1kin-minor}
show the major- and minor-axis stellar kinematics from the FORS1
spectra, and Figure~\ref{fig:n307-virus-kinematics} shows the kinematic
maps for the VIRUS-W data. Figure~\ref{fig:n307-compare-kinematics}
compares stellar kinematics extracted along the major axis from our
three datasets. Given the differences in the resolution for the
different observations (the respective FWHM or fiber sizes are indicated
by vertical shaded regions in the figures) -- and the relative noisiness
of the higher-order $h_{3}$ and $h_{4}$ moments -- the overall agreement
between the three datasets is good.

Both the major- and minor-axis long-slit kinematics show that the
velocity dispersion rises quite steeply in the inner $r \la 5\arcsec$,
suggestive of a kinematically hot central component. This can also be
seen, less clearly, in the higher dispersion of the central three bins
of the VIRUS-W data. There is, non the less, evidence for significant
rotation as in this region as well, as can be seen in the strong inner
velocity peak at $r \sim 3\arcsec$ and accompanying $V$--$h_{3}$
anti-correlation in the major-axis kinematics
(Figure~\ref{fig:n307-fors1kin-major}). Outside this region, the
kinematics are strongly rotation-dominated, with an observed peak
velocity of $\sim 200$~\kms{} and a dispersion profile that declines to
below 100~\kms{} for $r \ga 20\arcsec$ along the major axis.

As a whole, then, the stellar kinematics suggest a kinematically hot
central region (e.g., a classical bulge, albeit one with significant
rotation, or possibly a fast-rotating subcomponent) within the central
5\arcsec{} and a dominant disc component at larger radii. As we will
show below, this is consistent with both our stellar population analysis
of the FORS1 spectra and with our morphological analysis and 2D
decomposition of the galaxy.

There is photometric evidence for a weak bar or lens in NGC~307,
extending to about 10\arcsec{} in radius (see Sections~\ref{sec:n307}
and \ref{sec:2dfits}). Is there any evidence for this bar in the stellar
kinematics? We compare the observed kinematics with predictions from
$N$-body models published by \citet{bureau05} and \citet{iannuzzi15},
paying particular attention to projections where the bar orientation is
similar to that in NGC~307 (i.e., with the bar viewed nearly side-on).
Although some of the $N$-body model projections show a ``double-hump''
major-axis velocity profile, which might seem to agree with the clear
double-peak in NGC~307's velocity profile (upper left panel of
Figure~\ref{fig:n307-fors1kin-major}), this feature is only visible in
the models when the bar is close to \textit{end}-on, and vanishes when
the bar is closer to side-on. The double-hump velocity feature in
NGC~307 is thus almost certainly not a bar signature; it is more likely
due to a rapidly rotating substructure within the classical bulge
region.

The models do predict local extrema in $h_{3}$ -- and maxima in
$h_{4}$ -- near the ends of a strong bar seen side-on and at
inclinations of $75$ or 80\degr{} (e.g., lower right subpanels of Fig.~4
in \citealt{bureau05}). While there are local extrema in NGC~307's
$h_{3}$ profile at $r \sim 6$\arcsec which might be consistent with this
prediction, there are no such features in the $h_{4}$ profile
(Figure~\ref{fig:n307-fors1kin-major}). We conclude that
there is no evidence that the bar/lens strongly affects the stellar
kinematics in this galaxy.

\subsection{Quadrants for Stellar Kinematics}

As noted above (Section~\ref{sec:sinfoni-kinematics}), the SINFONI
kinematics were derived using a radial-angular binning scheme, with the
galaxy divided into four quadrants whose boundaries were the major and
minor axes of the galaxy. Each quadrant was subdivided into five angular
bins of varying width, with seven radial bins spaced logarithmically out
to the edge of the SINFONI field of view
(Figure~\ref{fig:sinfoni-kinematics}). To include the optical kinematics
in the same scheme for our dynamical modeling, we extended the quadrants
with additional radial bins and assigned values from the optical
kinematics. 

Since the FORS1 long-slit orientations were along the quadrant
boundaries, we assigned their kinematic values to the corresponding bins
along the quadrant boundaries -- e.g., the major-axis data were assigned
to corresponding closest bins along the major-axis boundaries of the
quadrants. For the Voronoi-binned VIRUS-W kinematics, we assigned each
Voronoi bin's kinematic values to the radial-angular bin containing the
center of the Voronoi bin.

\section{Stellar Population Analysis}\label{sec:stellar-pop}

To get a preliminary sense of how stellar populations -- and thus $M/L$ ratios --
might vary within NGC~307, we performed a stellar-population analysis of our
FORS1 long-slit spectroscopy.

We measured the Lick line strength index profiles from H$\beta$ to
Fe5406 as in \citet{mehlert00}. Following the minimum \chisquare{}
procedure described in \citet{saglia10}, we determined the age,
metallicity, and [$\alpha/$Fe] overabundance profiles that best
reproduced the observed profiles of the Lick indices H$\beta$, Mg$b$,
Fe5015, Fe5270, Fe5335, and Fe5406 using the simple stellar population
(SSP) models of \citet{maraston98,maraston05}, with a \citet{kroupa01}
IMF and the modeling of the Lick indices with $\alpha$-element
overabundance of \citet{thomas03a}.  We are able to reproduce the Mg and
Fe indices quite well; however, the measured H$\beta$ is systematically
$\approx 0.2$ \AA{} smaller than the models. As a consequence of this,
the resulting ages hit the maximum allowed value (15 Gyr) of the
model grid for most of the cases.  The [$\alpha/$Fe] profile is
approximately flat at a level of +0.3 dex, on both the major and minor
axes.

\begin{figure*}
\begin{center}
\includegraphics[scale=0.85]{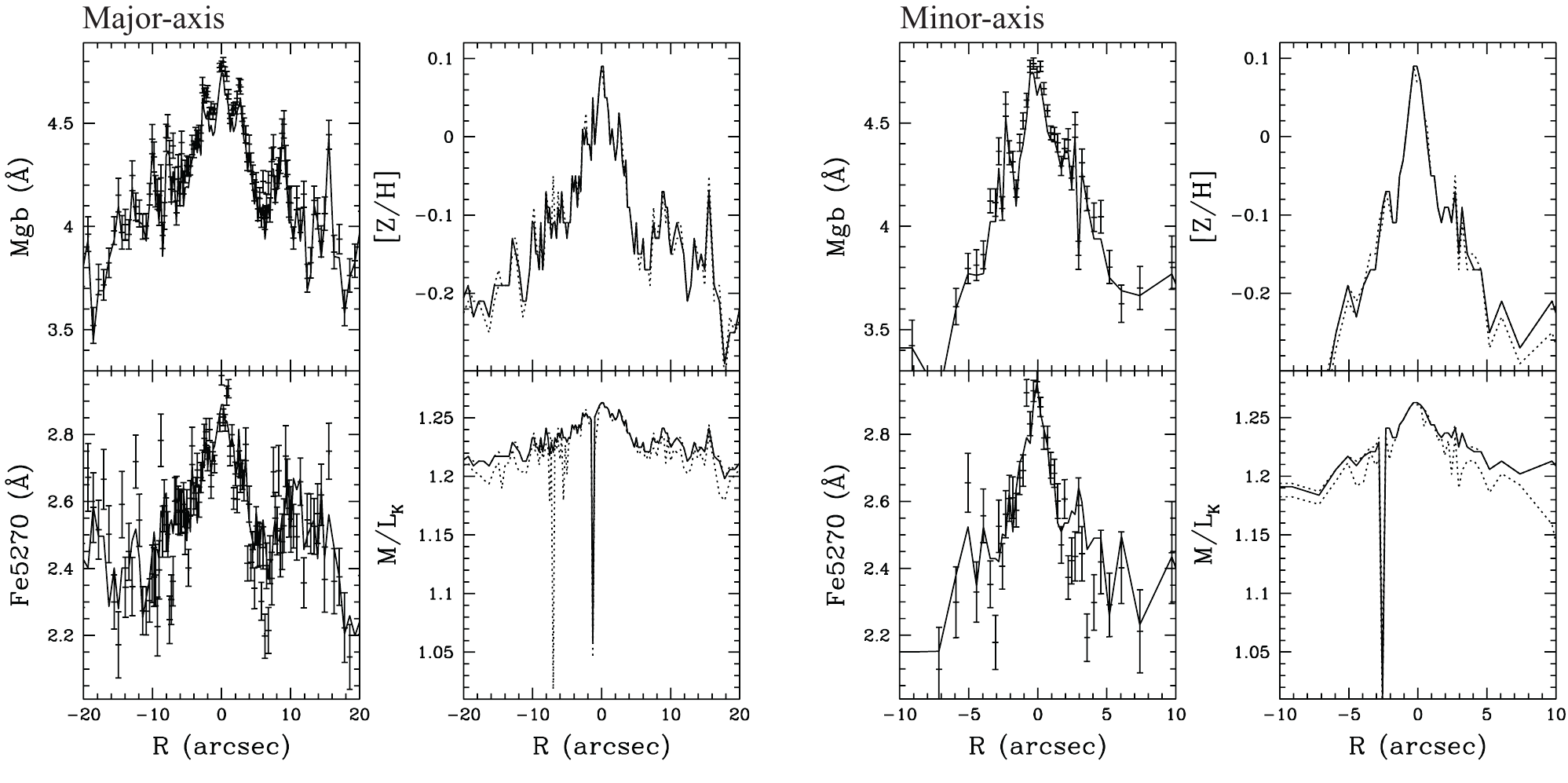}
\end{center}

\caption{Stellar-population analysis of VLT-FORS1 long-slit spectrum of NGC~307,
showing major-axis (left panels) and minor-axis (right panels) results.
In each set of panels, the left-hand two panels show examples of measured absorption-line indices (top: Mgb;
bottom: Fe5270); the right-hand two panels show results of the analysis (top: best-fitting
SSP metallicity; bottom: best-fitting SSP $K$-band $M/L$ ratio,
assuming a Kroupa IMF). Both major- and minor-axis profiles show a strong central
increase in stellar metallicity, and a weaker increase in the $M/L$ ratio. We
associate both with the dominance of a distinct ``classical bulge'' component
in the inner $r \la 5\arcsec$, also seen in the stellar kinematics and morphology.
\label{fig:n307-stellar-pops}}

\end{figure*}

Figure~\ref{fig:n307-stellar-pops} show some of the results, including
both raw Mgb and Fe5270 index measurements and the overall metallicity
([Z/H]) and $K$-band stellar $M/L$ ratio estimates.  The metallicity is
slightly above the solar value in the inner $r \la 5\arcsec$ and drops
to half-solar outside. The $K$-band $M/L$ ratio implied by the derived
age and metallicity profiles is approximately constant at a value of
1.22 $M_\odot/L_\odot$ at radii $\ga 10\arcsec$, rising to a central
peak of $\sim 1.26$. Actual radial variations in the $M/L$ ratio are
probably underestimated due to the saturated SSP age estimates.

Both major- and minor-axis profiles show evidence for a central
peak in metallicity, with a correspondingly higher $M/L$ ratio. This is
good evidence for a separate, metal-rich population with a higher
$M/L$ ratio dominating the inner $r \la 5\arcsec$ along the major
axis. As noted above, our VLT-FORS1 kinematics
(Figure~\ref{fig:n307-fors1kin-major} and \ref{fig:n307-fors1kin-minor})
show that the stellar velocity dispersion increases rapidly towards the
centre in this same region, from a nearly constant disc value of $\sim
110$--120 \kms{} to values $> 200$ \kms, suggesting a classical,
dispersion-dominated (albeit rapidly rotating) bulge. This is also
consistent with the decompositions we perform (below), which argue for a
relatively round luminosity component dominating the light at $r \la 5\arcsec$,
and motivates separating out the bulge component and allowing it to have
its own $M/L$ ratio in the modeling process.

\section{Photometric Modeling}

\subsection{General Approaches}

To properly measure the mass of a galaxy's SMBH, we must construct a
dynamical model based on at least two components: the potential of the
central SMBH and the potential due to the stellar mass distribution. (In
some cases, gas may also form a significant component; however, in
\citealt{mazzalay13}, we presented evidence that the molecular gas
content in the centres of the disc galaxies we observed with SINFONI --
that is, those galaxies where we \textit{could} detect gas -- was much
lower than the stellar mass in the same region, and so could reasonably be
neglected. In the case of NGC~307, we detected no gas emission at all;
an absence of significant gas is consistent with its S0 classification
and the lack of visible dust lanes in the optical images.)
The stellar-mass potential is the combination of a stellar 
$M/L$ ratio -- something adjusted during the fitting process -- and a
luminosity-density model for the stellar light. The luminosity-density
model, in turn, is derived from the observed stellar light distribution
of the galaxy, usually by deprojecting an observed surface-brightness
model. In this section, we describe how we devise luminosity-density
models for NGC~307.

The standard approach for constructing luminosity-density models has
been to fit ellipses to the isophotes of a galaxy image, and use the
resulting ellipse-fit model -- i.e., surface brightness, ellipticity,
and possibly symmetric higher-order terms ($\cos 4 \theta$, $\cos 6
\theta$, etc.) as a function of semimajor axis -- as input to the code
which then deprojects this to obtain a 3D luminosity-density model. An
alternate approach is to model the isophotes as the sum of multiple 2D
Gaussians, which can then be deprojected individually and summed to form
the luminosity density model \citep{emsellem94,cappellari02}. In the
case of something simple like most elliptical galaxies, this is usually
a straightforward operation, since we can assume that the entire galaxy
is a single, coherent stellar component.

But in constructing photometric models of disc galaxies, we face two
problems. The first has to do with questions of stellar $M/L$ ratios. As
noted above, most Schwarzschild modeling in the past has assumed a
single $M/L$ ratio for the entire stellar component. While this is
perhaps reasonable for elliptical galaxies, disc galaxies are known to
contain multiple stellar populations which can dominate different
regions of a galaxy. In the simplest case, a disc galaxy may have
distinct populations belonging to the bulge and to the disc; our
spectroscopic analysis suggests this is indeed the case for NGC~307
(Section~\ref{sec:stellar-pop}).

One possible approach is to consider a $M/L$ ratio which varies as a
simple function of radius \citep{mcconnell13}. But this may or may not
have a plausible physical origin, and there are a potentially unlimited
number of possible radial profiles to choose from, with varying numbers
of additional free parameters; even a linear function adds two extra
free parameters to the modeling process. We choose instead a somewhat more
physically motivated approach: we assume that the galaxy can be
spatially decomposed into two or more overlapping but distinct stellar
components, each with its own $M/L$ ratio \citep{davies06,
nowak10,rusli11}.

The second problem we have when constructing photometric models stems
from the fact that our Schwarzschild modeling code assumes an
\textit{axisymmetric} stellar potential, which can be described as a set
of coplanar, axisymmetric spheroids with relative thicknesses which can
vary as a function of radius (i.e., spheroids with $a = b$ but varying
vertical scale heights $c$). This requires an axisymmetric photometric
model as input to the deprojection algorithm: the isophote
\textit{shapes} can vary (in ellipticity and higher-order moments), but
their \textit{orientations} (position angles) cannot. Real disc
galaxies, however, are often non-axisymmetric, with bars, spiral arms,
and other stellar substructure which show up in ellipse fits as
variations in ellipticity \textit{and} position angle. Since the
deprojection process cannot handle position-angle variations, they are
ignored, and the result is that changes in isophotal ellipticity due to,
e.g., bars or spiral arms are misleadingly converted into changes in
vertical thickness in the resulting luminosity-density model.

To deal with these issues, we use an approach first described in
\citet{nowak10} and applied to the galaxies NGC~3368 and NGC~3489 in
that paper, and also to NGC~1332 in \citet{rusli11}. This consists of
first identifying plausible ``bulge'' and ``disc'' regions, devising
preliminary models corresponding to the bulge and disc, creating
separate residual images for the two components (i.e., a ``bulge-only''
image which has the disc model subtracted off and a ``disc-only''
image with the bulge model subtracted off), and then treating them
in distinct fashions: 

\begin{enumerate} 

\item The bulge-only residual image is fit with freely varying ellipses
in the standard fashion, treating it as though it were the image of a
spheroidal, axisymmetric structure with potentially variable $c/a$ axis
ratios.

\item The disc-only residual image is fit with ellipses which are
\textit{fixed} to a common shape and orientation (axis ratio and
position angle) corresponding to that of the outer disc. This has the
effect of \textit{azimuthally averaging} whatever non-axisymmetric
structure -- bars, spiral arms, etc. -- may actually exist.
\end{enumerate}

Although we generate preliminary models for both bulge and disc based on
combinations of simple analytic components (e.g., an elliptical S\'ersic
component for the bulge), the final surface-brightness models which we
pass to the deprojection machinery are based primarily on direct ellipse
fits to the residual images as outlined above. This means that the final
models -- especially for the bulge component -- contain as much of the
intrinsic galaxy light variation as possible: e.g., our final bulge
component is not a pure S\'ersic component, but represents the galaxy
light after the preliminary disc model has been subtracted.

In the specific case of NGC~307, as we will discuss below, there is
evidence for a rounder stellar ``halo'' which dominates the light beyond
a certain radius. Thus, we modify the second surface-brightness
component described above by allowing the isophotes to have lower
ellipticities (as measured by ellipse fits with variable ellipticity) at
large radii.

\subsection{Photometric Modeling of NGC~307}\label{sec:2dfits} 

As noted above, there is evidence for a central bulge in NGC~307 with a
distinct metal-rich stellar population dominating the inner $r \la
5\arcsec$, a weak bar or lens contributing to the light at intermediate
radii (most strongly at $r \sim 9$--10\arcsec), and a halo dominating
the outer light ($r \ga 50\arcsec$). Therefore, we analysed this galaxy
with a 2D decomposition approach, including up to four components:
central bulge, bar/lens, disc, and halo. (Note that in this subsection
we use ``halo'' to refer specifically to a \textit{stellar} component,
not to a dark-matter halo.)

To start with, we fit the FORS1 image with \textsc{Imfit}
\citep{erwin15-imfit} using several models, including both a simple
bulge + disc (B+D = S\'ersic + exponential) model and two versions of a
bulge + bar/lens + disc (B+b+D) model, which differed in how the
bar/lens was modeled. A Moffat PSF based on the median values of fits to
stars in the image was convolved with each model during the fitting
process.  We compared the effectiveness of the models using the Akaike
Information Criterion \citep[AIC;][]{akaike74}, which is automatically
computed by \textsc{Imfit} based on the likelihood of the best-fitting
model, the number of data points, and the number of
parameters.\footnote{\textsc{Imfit} actually computes the ``corrected''
version of AIC (AIC$_{\rm c}$), though given the large number of
individual data points involved, the difference between AIC$_{\rm c}$
and AIC is minimal.} Lower values of AIC indicated (relatively) better
fits. A difference in AIC values between two models of $< 2$ is
considered insignificant, while a difference $> 6$ is considered strong
evidence for the model with lower AIC being better.

The best of these models, with the lowest AIC, was the B+b+D model with
the bar/lens represented by an elliptical, broken-exponential component
\citep{erwin08,erwin15-imfit}. Using a S\'ersic function for the
bar/lens provided a reasonable fit, though not nearly as good
($\Delta$AIC = +675 relative to the broken-exponential model). The
baseline B+D model was a much poorer fit, with $\Delta$AIC = +4291
relative to the broken-exponential model.

To determine the contribution of the halo component, we then performed a
four-component (B+b+D+H) fit to the (larger FOV) WFI image, starting
with the best B+b+D model from the FORS1 image fits and adding a
S\'ersic component with generalized ellipses (i.e., boxy or discy
isophote shapes) to represent the halo. Generalized ellipses are described
by 
\begin{equation}
\left( \frac{|x|}{a} \right)^{c_{0} \,+\, 2} \!\! + \; \left( \frac{|y|}{b} \right)^{c_{0} \,+\, 2}  = \; 1,
\end{equation}
where $|x|$ and $|y|$ are distances from the ellipse centre in the
coordinate system aligned with the ellipse major axis, $a$ and $b$ are
the semi-major and semi-minor axes, and $c_{0}$ describes the shape:
$c_{0} < 0$ corresponds to disky isophotes, $c_{0} > 0$ to boxy
isophotes, and $c_{0} = 0$ for perfect ellipses. The best-fitting halo
component had slightly boxy isophotes ($c_{0} = 0.57$) and a profile
essentially indistinguishable from an exponential (S\'ersic $n = 0.97$);
this component is slightly misaligned with respect to the disc and bulge
(both disc and bulge have PA $\approx 82\degr$, while the halo has PA
$\approx 77\degr$). We then re-fit the (higher-resolution) FORS1 image
by including the halo component, keeping most of its structural
parameters fixed to the best-fitting values from the WFI fit but allowing
the position angle and intensity ($I_{e}$) to be free parameters. 

\begin{figure*}
\begin{center}
\includegraphics[scale=0.95]{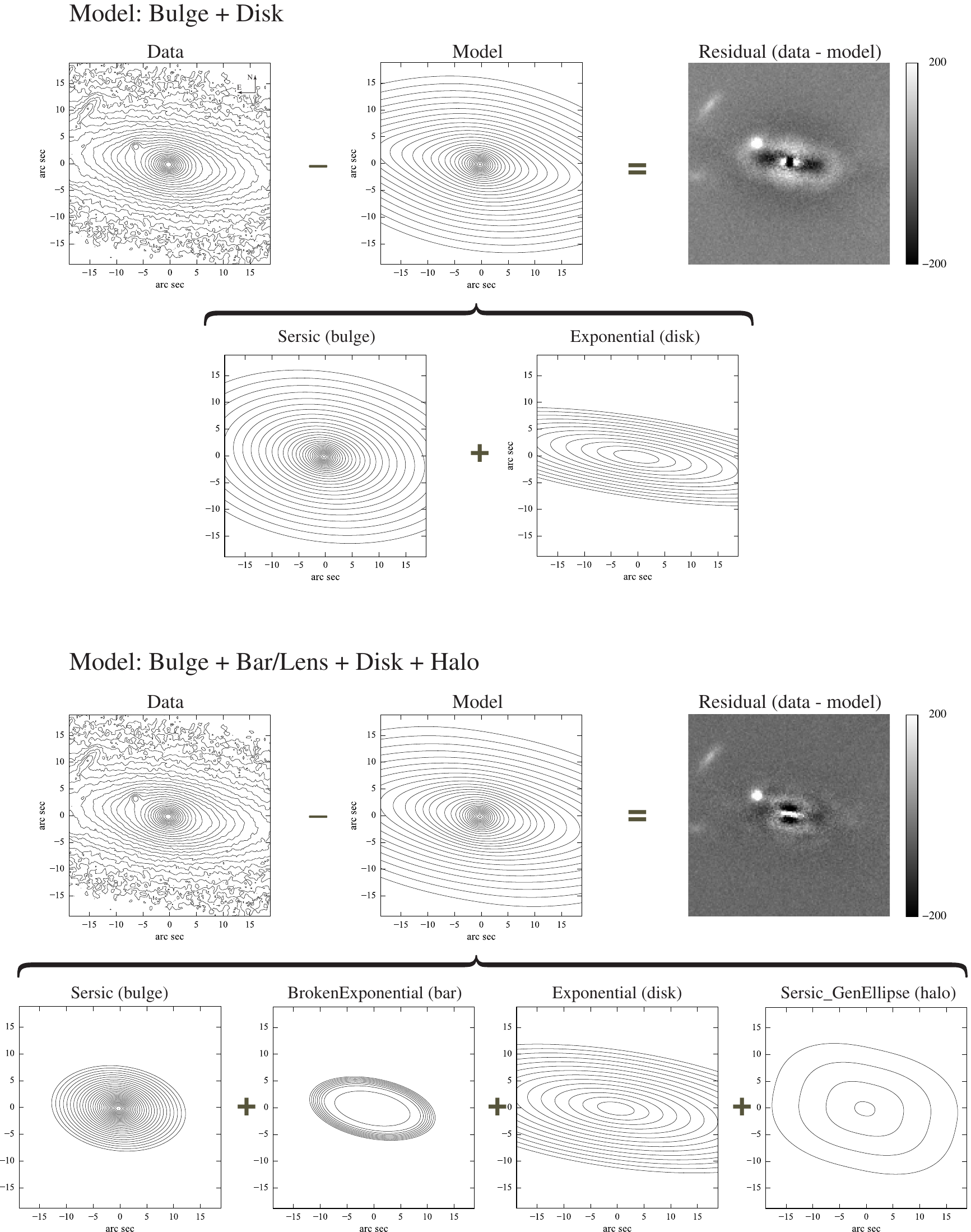}
\end{center}

\caption{Comparison of different two-dimensional decompositions of
NGC~307. For each model, we show in the first row the observed,
logarithmically scaled data isophotes (VLT-FORS1 $R$-band, left), model
isophotes (middle), and the residual image (data $-$ model, linear
scaling from $-200$ to 200 counts/pixel; right). The individual
components contributing to the model are show in the second row;
components are identified by the name of the \Imfit{} function used for
each \citep[see][]{erwin15-imfit}. Top pair of rows: best-fitting simple
bulge + disc model. Bottom pair of rows: best-fitting bulge + bar/lens +
disc + outer halo model. The range of isophote levels is the same for
all contour plots, and both residual images use the same display range.
North is up and east is to the left in all panels.
\label{fig:n307-decomp}}

\end{figure*}

\begin{figure}
\begin{center}
\hspace*{-12.5mm}
\includegraphics[scale=0.59]{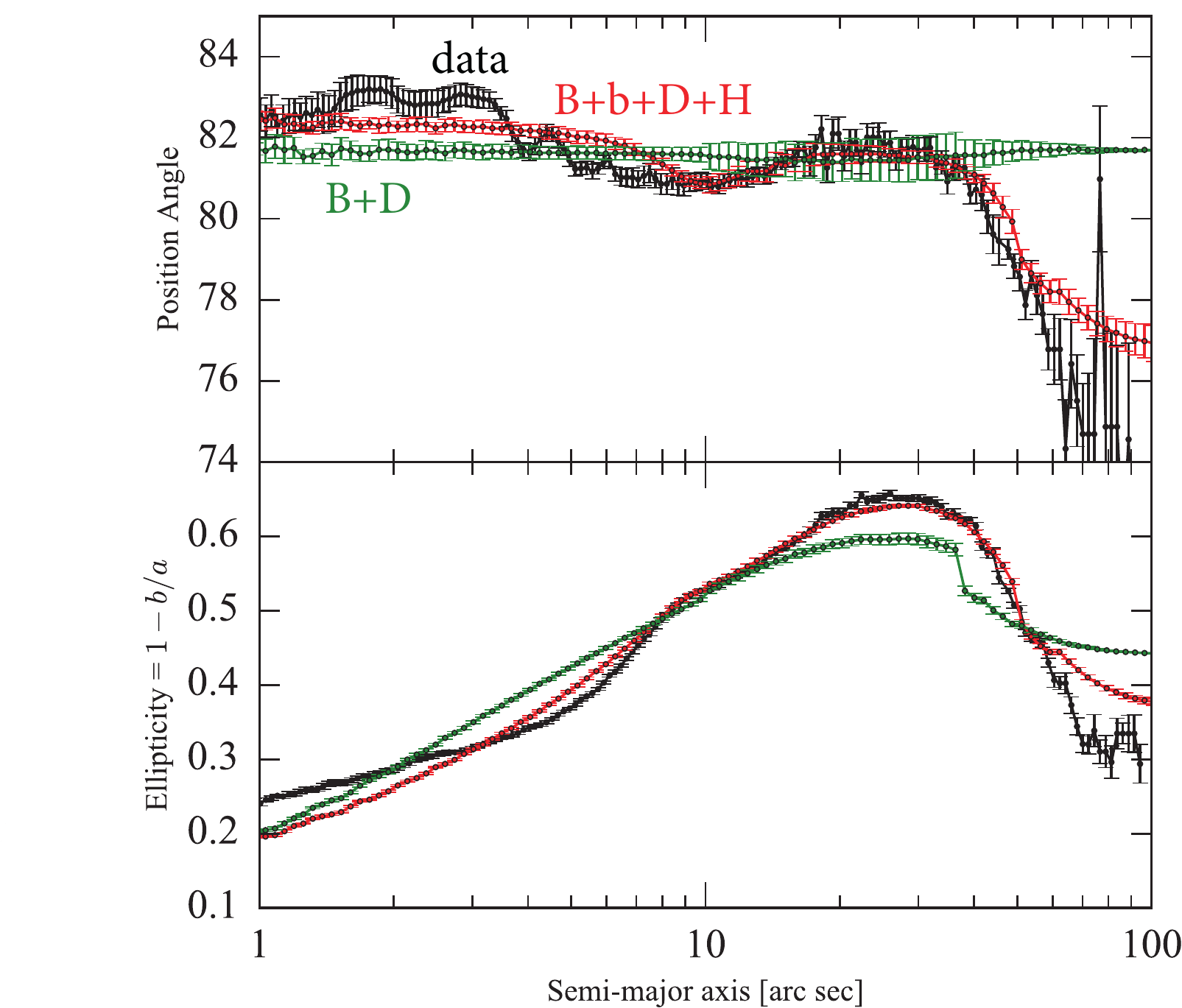}
\end{center}

\caption{Position angles and ellipticities from ellipse fits to the
VLT-FORS1 and WFI images of NGC~307 (black, using FORS1 data for $a <
46\arcsec$), the best-fitting, PSF-convolved 2D bulge + disc model image (B+D, green),
and the best-fitting, PSF-convolved 2D bulge + bar + disc + outer halo model image
(B+b+D+H, red). The latter model is significantly better at reproducing
the isophote shapes. \label{fig:n307-efit-data-vs-model}}

\end{figure}

Figure~\ref{fig:n307-decomp} compares our final four-component B+b+D+H
fit to the FORS1 image (lower panels) with the baseline B+D fit (upper
panels); the parameters of the B+b+D+H fit are listed in
Table~\ref{tab:n307-decomp}. In addition to the fact that the second
decomposition is a significantly better fit in a statistical sense
(e.g., $\Delta$AIC is $-5486$ relative to the B+D model, and $-1491$
relative to the best B+b+D model), we can see that the B+D fit has an
exceptionally narrow disc (ellipticity = 0.80) and an exceptionally
bright bulge component with S\'ersic index $n = 5.5$; the value of $n =
2.5$ for the bulge in the B+b+D+H fit is much more typical of bulges in
S0 galaxies \citep{laurikainen10}.
Figure~\ref{fig:n307-efit-data-vs-model} compares ellipse fits to the
data (black) and to the B+D (green) and B+b+D+H (red) model images; the
latter does a significantly better (albeit not perfect) job of matching
position-angle twists and ellipticity variations in the data.

Figure~\ref{fig:n307-major-axis} shows the galaxy's major-axis
surface-brightness profile from the FORS1 image, along with major-axis
cuts through the PSF-convolved B+b+D+H model (dashed black line) and the
individual components of the model. This shows that the inner S\'ersic
component dominates the light for $r \la 5\arcsec$ -- making it a very
plausible match to the separate stellar population suggested by our
spectroscopic analysis (Section~\ref{sec:stellar-pop}). We note that the
ellipticity of this component (0.385) is a good match to the observed
outer isophote ellipticity in the SINFONI image ($\sim 0.4$), where
seeing effects are smallest.

\subsubsection{Generating Final ``Bulge'' and ``Disc'' Components for Dynamical Modeling}
\label{sec:bulge-disk-comps}

To generate ``bulge-only'' images for use in constructing the final
bulge model, we constructed model images (using the \texttt{makeimage}
tool in \Imfit) consisting of the bar/lens, disc, and halo components of
the best-fitting B+b+D+H model, suitably rescaled and PSF-convolved for
the SINFONI, FORS1, and WFI images. These were then subtracted from the
data images, and ellipses were fit to the resulting residual
images. The final bulge profile consisted of ellipse-fit data from the
residual SINFONI image for $a < 1.1\arcsec$, FORS1 data for $a =
1.1$--10\arcsec, and a S\'ersic extrapolation of the inner data (using
the bulge parameters in Table~\ref{tab:n307-decomp}) for larger 
radii.\footnote{The surface brightness of the residual bulge image is too low
and noisy to be fit outside $a \sim 10\arcsec$; this is also true if we
use the WFI image.}
The ellipticity and $\cos 4 \theta$ values were taken from the residual
SINFONI image ellipse fits for $a < 1.1\arcsec$ and were set to 0.385
and 0, respectively, for larger radii.

The ``disc-only'' images used for constructing the final disc model
(actually the disc + lens + stellar halo) were generated in an analogous
fashion: PSF-convolved model-bulge images (using the inner S\'ersic
parameters from Table~\ref{tab:n307-decomp}) were subtracted from the
FORS1 and WFI images, and the resulting residual images were fit with
both fixed and free ellipses. Since, as noted above, NGC~307 has a
significant outer halo which is rounder than the disc, the final
``disc'' model actually incorporates a transition from the azimuthally
averaged, constant-ellipticity profile  to a profile with declining
ellipticity at $a = 28\arcsec$. The ellipticity and PA for the
fixed-ellipse fits were 0.69 and 82\degr, based on free-ellipse fits to
the residual FORS1 image; these values are almost identical to those of
the exponential-disc component in the best-fitting B+b+D+H model
(Table~\ref{tab:n307-decomp}). The fixed-ellipse-fit FORS1 data were
used for $a = 6.4$ to 28\arcsec{} in the final profile, with
free-ellipse-fit surface-brightness and ellipticity used for $a \ge
28\arcsec$ (using WFI free-ellipse-fit data for $a > 47\arcsec$). For $a
< 6.5\arcsec$, the FORS1 fixed-ellipse-fit surface brightness became
extremely noisy and difficult to deproject; thus the surface-brightness
data at smaller radii come from a fixed-ellipse fit to an unconvolved
model image (built using the disc + bar/lens + halo components from
Table~\ref{tab:n307-decomp}). At these small radii the final luminosity
density is dominated by the bulge component, so accuracy in the disc
component is less important.

Figure~\ref{fig:n307-sb} shows the surface-brightness profiles of the
final bulge and disc components. The top panel of
Figure~\ref{fig:n307-sb-comp} compares the surface-brightness profile of
our final bulge component (red) with the equivalent
(ellipse-fit-derived) surface-brightness profile of the S\'ersic
function (convolved with the SINFONI PSF) from our 2D decomposition.
Although they are very similar, the final bulge component is brighter in
the centre than the inward extrapolation of the S\'ersic function. If we
had simply used the S\'ersic function itself as the bulge component for
deprojection, we would underestimate the central stellar density and
thus potentially overestimate the SMBH mass in our modeling. The bottom
panel of the figure makes the same comparison for the disc component.

\begin{figure}
\begin{center}
\hspace*{-1.5mm}
\includegraphics[scale=0.45]{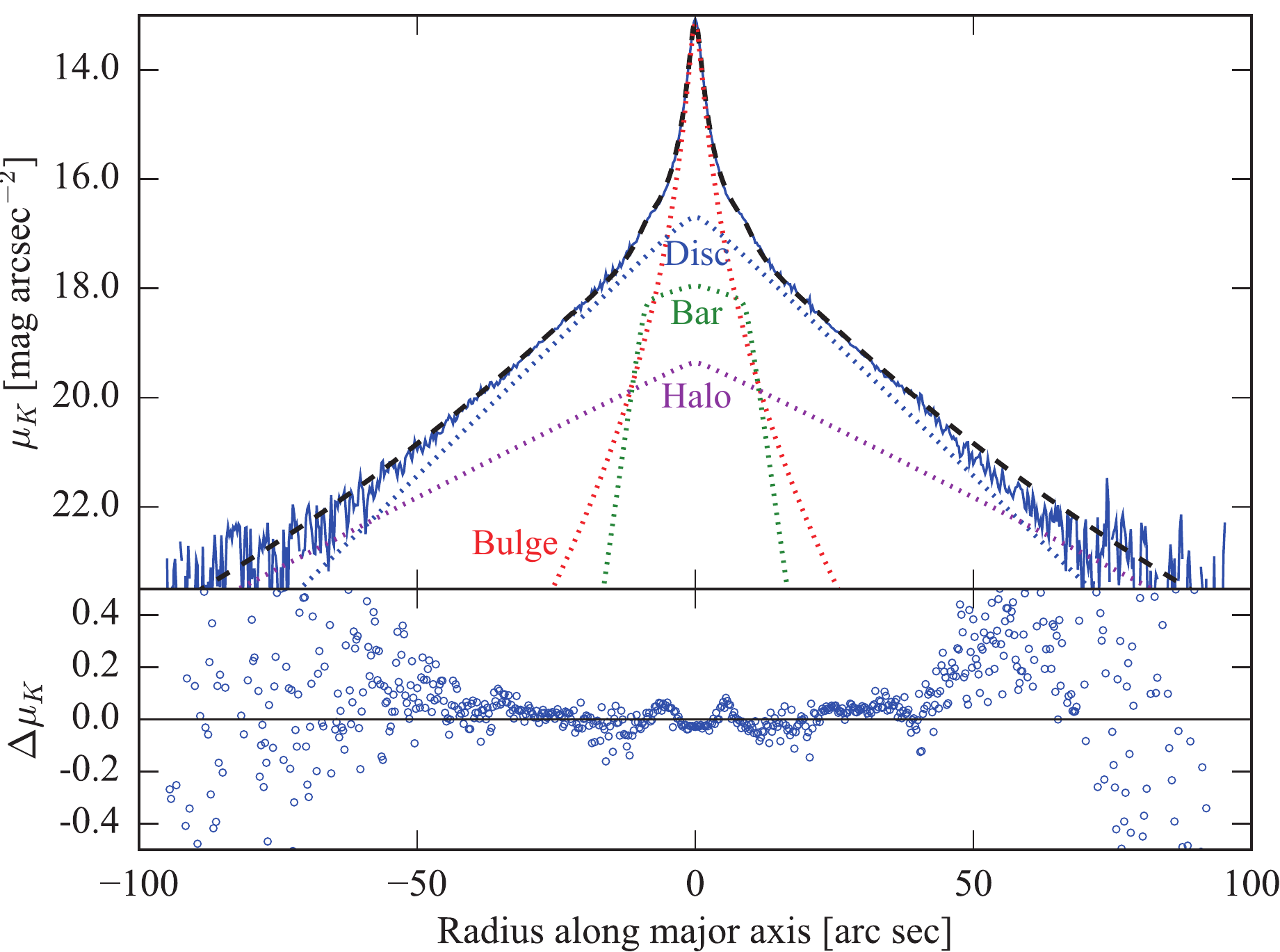}
\end{center}

\caption{Upper panel: Major-axis cut through the VLT-FORS1 and WFI
$R$-band images of NGC~307 (solid blue line, using FORS1 data for $|r| <
22\arcsec$), along with major-axis cuts through best-fitting,
PSF-convolved B+b+D+H model image (black dashed line) and through individual
(PSF-convolved) components of the model image: S\'ersic (bulge; red
short-dashed line), broken-exponential (bar/lens; green short-dashed
line), exponential (disc; blue short-dashed line), and outer S\'ersic
(halo; magenta short-dashed line). Lower panel: Residuals from fit
($\mu_{\rm data} - \mu_{\rm model}$) evaluated along the major-axis
cut.\label{fig:n307-major-axis}}

\end{figure}

\begin{figure}
\begin{center}
\hspace*{-5.5mm}
\includegraphics[scale=0.6]{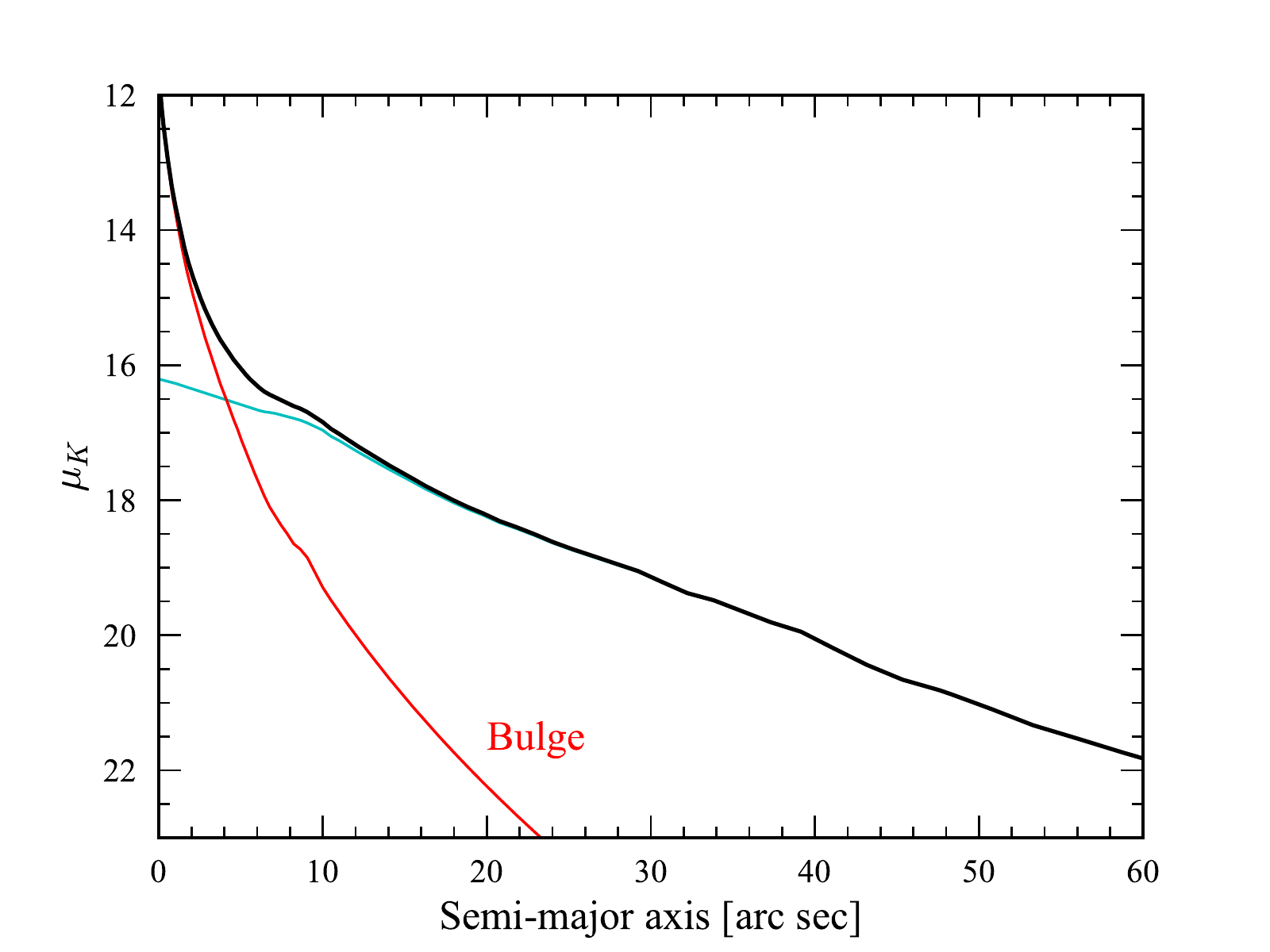}
\end{center}

\caption{Final surface-brightness profiles used for constructing stellar
luminosity-density components for NGC~307, showing $K$-band surface
brightness versus semi-major axis for the ``disc'' (i.e., bar + disc +
halo; cyan) and bulge (red) components, as well as their sum (thicker black line).
\label{fig:n307-sb}}

\end{figure}

\begin{figure}
\begin{center}
\hspace*{-3mm}
\includegraphics[scale=0.88]{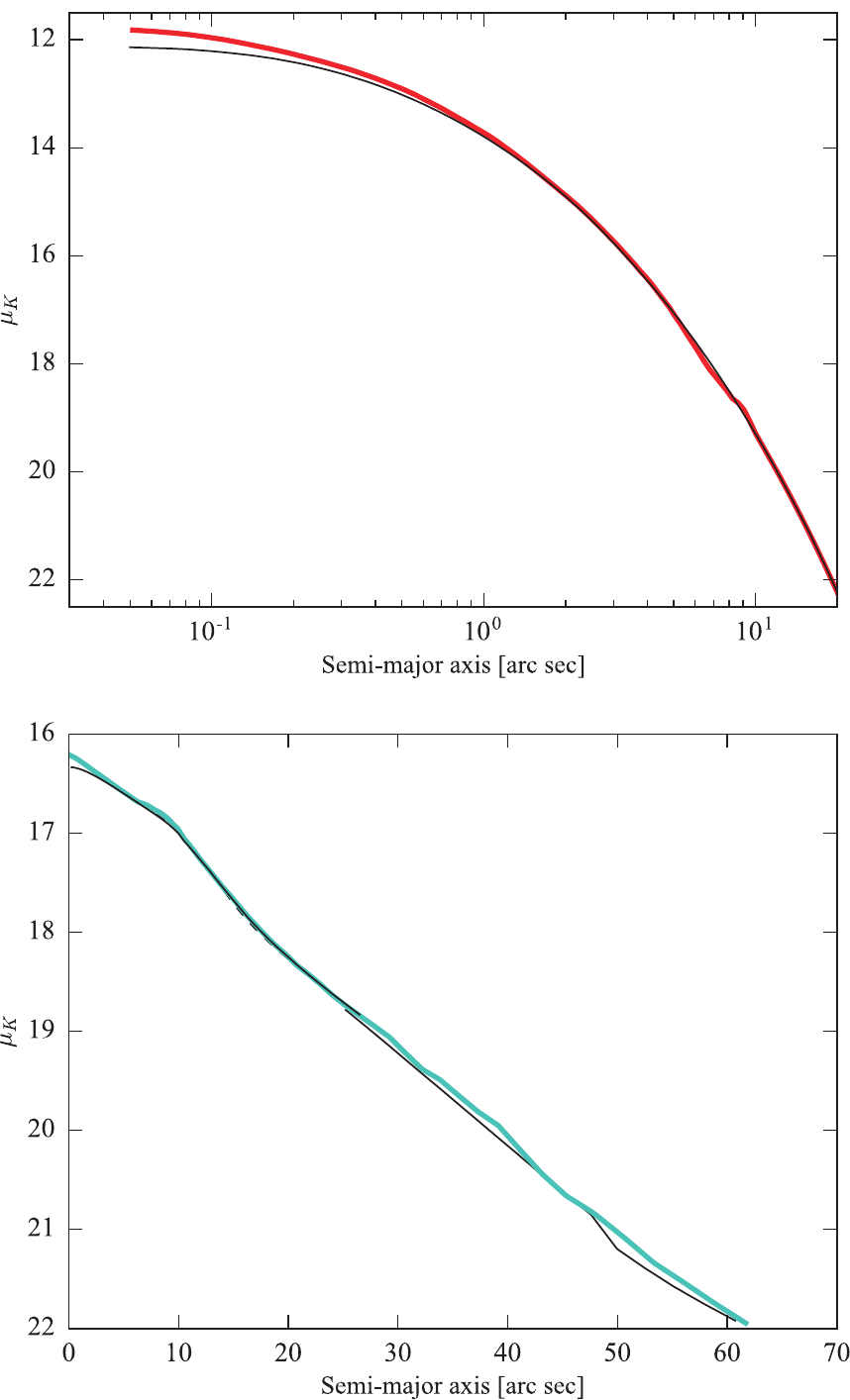}
\end{center}

\caption{Top: Comparison of the final surface-brightness profile for the
measured bulge component, used to construct the bulge luminosity-density
component for our dynamical modeling (thick red line, same as in
Figure~\ref{fig:n307-sb}), with the profile of the S\'ersic component
from our 2D decomposition (thin black line, convolved with the
SINFONI PSF). Bottom: Comparison of the final disc component (thick cyan
line, same as in Figure~\ref{fig:n307-sb}) with the profile of the
exponential + broken-exponential + outer S\'ersic components from our 2D
decomposition (thin dashed black line = unconvolved model image,
thin solid black line = convolved with VLT-FORS1 PSF); values at $a
< 28\arcsec$ are from fixed ellipse fits, while values at larger
semi-major axis values are from free ellipse fits (see text).
\label{fig:n307-sb-comp}}

\end{figure}

Ideally, one could treat the outer halo as a third stellar component,
with its own $M/L$ ratio. However, since our kinematic data are limited
to $r \la 30\arcsec$ along the major axis, well inside the region where
the halo component begins to dominate over the disc (e.g.,
Figure~\ref{fig:n307-major-axis}), the precise details of the stellar
halo do not significantly affect our dynamical modeling.

\begin{table}
\caption{NGC 307: 2D Photometric Decomposition}
\label{tab:n307-decomp}
\begin{tabular}{@{}llrrl}
\hline
Component          & Parameter & Value & $\sigma$  & Units \\
\hline
Sersic             &  PA          & 82.41  & 0.18   &  deg \\
(bulge)            &  $\epsilon$  & 0.385  & 0.002  &  \\
                   &  $n$         & 2.548  & 0.041  &  \\
                   &  $\mu_{e}$   & 15.076 & 0.035  &  mag arcsec$^{-2}$\\
                   &  $r_{e}$     & 2.186  & 0.049  &  arcsec \\
BrokenExponential  &  PA          & 79.19  & 0.29  &  deg \\
(bar/lens)         &  $\epsilon$  & 0.552  & 0.005  &  \\
                   &  $\mu_{0}$   & 17.907 & 0.170  &  mag arcsec$^{-2}$\\
                   &  $h_{1}$     & 32.39  & 18.61  &  arcsec \\
                   &  $h_{2}$     & 1.521  & 0.095  &  arcsec \\
                   &  $R_{\rm brk}$     & 9.090 &  0.105  &  arcsec \\
                   &  $\alpha$    & 10.0   & ---   &  arcsec$^{-1}$ \\
Exponential        &  PA          & 81.81  & 0.29  &  deg \\
(disc)             &  $\epsilon$  & 0.708  & 0.002  &  \\
                   &  $\mu_{0}$   & 16.509 & 0.017  &  mag arcsec$^{-2}$\\
                   &  $h$         & 11.04  & 0.08   &  arcsec \\
Sersic\_GenEllipse &  PA          & 76.80  & 0.90   &  deg \\
(halo)             &  $\epsilon$  & 0.349  & 0.003  &  \\
                   &  $c_{0}$     & 0.569  & 0.035  &  \\
                   &  $n$         & 0.972  & 0.016  &  \\
                   &  $\mu_{e}$   & 21.077 & 0.014  &  mag arcsec$^{-2}$\\
                   &  $r_{e}$     & 35.36  & 0.19   &  arcsec \\
\hline 
\end{tabular}

\medskip

Summary of the final 2D decomposition of the $R$-band VLT-FORS1 image of
NGC~307 (using \Imfit). Column 1: \Imfit{} component names. Column 2:
Parameter names. Column 3: Best-fit parameter value. Column 4: Nominal
uncertainty on parameter value (from Levenberg-Marquardt minimization).
Column 5: Units of the parameter. For all parameters of the
Sersic\_GenEllipse component except PA and $\mu_{e}$, the values come
from fitting the WFI image and were held fixed during the fit to the
FORS1 image. Surface brightnesses are in $K$-band.

\end{table}

\subsection{Deprojection} 
\label{sec:deprojection}

To go from the surface-brightness profiles and the accompanying
geometric information (ellipticity, $B_{4}$) to actual 3D luminosity
density models requires deprojection under certain assumptions. We use
an approach based on that of \citet{magorrian99}. Different realizations
of 3D luminosity-density models are projected, assuming an inclination
of 76\degr,\footnote{Based on the observed maximum ellipticity of
$\approx 0.69$ in the disc-dominated region, assuming an intrinsic disc
thickness of $c/a = 0.2$.} and compared to the observed 2D
surface-brightness model derived from the profiles. A simulated
annealing algorithm is used to maximize a penalized log-likelihood
function based on the difference between the model and the data in order
to determine the best-fitting 3D model. 

We performed separate deprojections for the bulge and disc components.
Since the central regions of the disc component are negligible compared
to the bulge component, we ignored the effects of PSF convolution (in
fact, as explained in the previous section, the central part of our disc
surface-brightness component was derived from an unconvolved model
image). For the bulge component, on the other hand, PSF convolution
\textit{is} important, so we used our double-Gaussian model of the
SINFONI PSF (Section~\ref{sec:sinfoni-data}) when projecting trial 3D
bulge-component models for comparison with the data.

\section{Dynamical Modeling}

To determine the SMBH mass and stellar $M/L$ ratios for NGC~307, we use
Schwarzschild orbit-superposition modeling \citep{schwarzschild79} with
the three-integral, axisymmetric code of \citet{thomas04}, which is
based in turn on the code of \citet{gebhardt03} (see also
\citealt{siopis09}).

The basic outline of our Schwarzschild modeling process is as follows.
First, we define general mass models consisting of a SMBH, one or more
stellar components, and (optionally) a DM halo. Then, for each such
model, we perform fits spanning a grid in the space of free parameters
of the model, computing regularized \chisquare{} values (see below) for
each combination of parameters. Finally, we analyse the resulting
\chisquare{} landscape and the corresponding likelihoods to determine
best-fit parameter values and corresponding confidence intervals.

The fitting process for a given general model is:
\begin{enumerate}

\item Construct a specific mass model and its potential from the general
model, based on particular values for the free parameters (SMBH mass,
stellar $M/L$ ratios, DM halo parameters).

\item Integrate test particles within this potential to build a library of
orbits. For NGC~307, we used $2 \times 14,300$ individual orbits, with
the duplication achieved by reversing the angular momentum of individual orbits.

\item Assign weights $w_{i}$ to the individual orbits so
that their weighted sum reproduces the input stellar mass model (this
is treated as a boundary condition, so the match is exact to
within machine tolerances\footnote{This helps ensure self-consistency,
so that the generated model reproduces the potential used to compute the
orbits.}) \textit{and} reproduces the observed kinematics. The
comparison with the kinematic data is done by simulating kinematic
observations of the model using the same spatial and LOSVD bins as the
data, convolved with PSFs based on the observations. A \chisquare{}
value is computed based on the comparison between the observed and model
kinematics.

\item Repeat the process with new values of the free parameters.
\end{enumerate}

The fit of a given orbit library to the kinematic data is computed by maximizing
$\hat{S} = S - \alpha \chisquare$. This is a regularized version of
a \chisquare{} minimization, based on a maximum entropy approach,
where $\alpha$ is the regularization parameter and $S$ is the Boltzmann entropy:
\begin{equation}
S \; = \; -\sum_{i} w_{i} \ln \left( \frac{w_{i}}{V_{i}} \right),
\end{equation}
with $V_{i}$ the phase-space volume of orbit $i$, computed as in \citet{thomas04}.
The \chisquare{} term is
\begin{equation}
\chisquare = \sum_{j = 1}^{N_L} \sum_{k}^{N_{\rm vel}} 
\frac{ (L_{jk,{\rm m}} - L_{jk,{\rm d}})^{2}}{\sigma_{jk}^{2}},
\end{equation}\label{eq:chi2}
which is a sum over the $N_{L}$ spatial positions $j$ and the $N_{\rm vel}$
LOSVD bins $k$, with $L_{jk,{\rm m}}$ and $L_{jk,{\rm d}}$ the model and data 
values in each LOSVD bin and $\sigma_{jk}^{2}$ the corresponding Gaussian 
uncertainty for the data. 

Since our modeling code assumes axisymmetry, we treat each quadrant of
kinematic data as a separate dataset to which the model is fit. The
result is four independent evaluations for each set of model parameters,
which can in principle be used as quasi-independent estimates of model
uncertainties, as well as a gauge of how well the underlying assumption
of axisymmetry is justified \citep[e.g.,][]{nowak10}. Our final analysis
is based on combining the results for all four quadrants, as described
below.

We have four general models. Each features a central
SMBH. \textbf{Model~A} has single stellar component; \textbf{Model~A+DM}
adds a DM halo to this. \textbf{Model~B} has \textit{two} stellar
components: one for the bulge sub-component and one for the
disc;\footnote{Where ``disc'' means the combined disc + bar/lens +
stellar halo component, as determined in Section~\ref{sec:bulge-disk-comps}.}
\textbf{Model~B+DM} also includes a DM halo. These models are summarized
in Table~\ref{tab:n307-models}, and described in more detail in the following
subsections.

The stellar density components are based on stellar \textit{luminosity
density} components $\nu$ (plus a $M/L$ ratio which converts luminosity
to mass). The luminosity density components themselves are obtained by
deprojecting the surface-brightness components
(Section~\ref{sec:deprojection}). For the single-stellar-component
model, we simply add the bulge and disc luminosity-density models
together and assign the result a single $M/L$ value. 

To determine best-fit values and confidence intervals for parameters, we
use a slightly modified version of the likelihood-based approach of
\citet{mcconnell11a} and \citet{rusli13a}.  For each
value of a given parameter (e.g., \mbh), we compute the relative
likelihood (from the \chisquare) for a given quadrant by marginalizing over the other
parameters; the final relative likelihood is then the product of the 
likelihoods for the individual quarters. As an example, the marginalized likelihood value
$\mathcal{L}_{n}(x)$ for a model with parameters $x$, $y$, and $z$, evaluated
in quadrant $n$, would be:
\begin{equation}
\mathcal{L}_{n}(x) \; \propto \; \sum_{y_{\rm min}}^{y_{\rm max}} 
\sum_{z_{\rm min}}^{z_{\rm max}} e^{-\frac{1}{2} \chi_{n}^{2}(x,y,z)} \Delta z \, \Delta y ,
\end{equation}
and the final marginalized likelihood value would be
\begin{equation}
\mathcal{L}(x) \; = \; \prod_{n = 1}^{4}  \mathcal{L}_{n}(x) .
\end{equation}

To determine the best-fit values and confidence intervals,
we use the cumulative of the marginalized likelihood:
\begin{equation}
C(x) \; = \;  \frac{ \int_{x_{\rm min}}^{x} \mathcal{L}(x^{\prime}) \,\mathrm{d}x^{\prime} }
{ \int_{x_{\rm min}}^{x_{\rm max}} \mathcal{L}(x^{\prime}) \,\mathrm{d}x^{\prime} }
\end{equation}
with the best-fit value at the median, where $C(x) = \frac{1}{2}$, and
the 68\% (``1-$\sigma$'') confidence interval defined by the values of
$x$ for which $C(x) = 0.16$ and 0.84.

The best-fit parameter values and confidence intervals for each model
are presented in Table~\ref{tab:n307-results}, along with the total CPU
time used for each model.\footnote{The code ran in a cluster with
approximately 500 Intel Xeon 2.6 GHz E5-2670 CPUs.} The relative
\chisquare{} and marginalized likelihood plots for SMBH mass and stellar
$M/L$ ratios for all four models are shown in
Figure~\ref{fig:n307-likelihood-all-models}. The grey shaded areas show
the (arbitrarily scaled) marginalized likelihood for the parameter in
question, with the best-fit value and confidence intervals indicated by
the solid and dashed vertical black lines. The lines show $\Delta
\chisquare = \chi^{2}(x) - \chi^{2}_{0}$, where $\chi^{2}(x)$ is the
minimum for all models with the same value of the parameter in question
(marginalized over the other parameters) and $\chi^{2}_{0}$ is the
minimum \chisquare{} over all parameter values. The thin lines show the
$\Delta \chisquare$ values for the individual-quadrant fits; the thick
lines are the result of summing the individual-quadrant \chisquare{}
values.

\begin{table}
\caption{NGC 307: Summary of Dynamical Models}
\label{tab:n307-models}
\begin{tabular}{@{}lllr}
\hline
Model Name    & Stellar Component(s)  & DM halo & $N_{\rm free}$ \\
(1)           & (2)                   & (3)      & (4) \\
\hline
A             & Single \mlt{}         & No       & 2 \\
A+DM          & Single \mlt{}         & Yes      & 4 \\
B             & \mlb, \mld{}          & No       & 3 \\
B+DM          & \mlb, \mld{}          & Yes      & 5 \\
\hline 
\end{tabular}

\medskip

Summaries of the different dynamical models fit to the kinematic data of
NGC~307. (1) Name of model. (2) Stellar component(s) = whether a single
stellar component (one $M/L$ ratio) or separate bulge and disc
components with independent $M/L$ ratios were used. (3) DM halo =
whether a dark-matter halo was used in the model. (4) The number of free
parameters in the model.

\end{table}

\subsection{Model A: SMBH + Single Stellar Component}
\label{sec:model-a}

Model~A is is the traditional model used for most published dynamical
SMBH mass measurements. It consists of a SMBH and a single
stellar-density component:
\begin{equation}
\rho \; = \; \mbh \, \delta(r) \: + \: \mlt \, \nu_{\rm tot}.
\end{equation}
For NGC~307, the single luminosity-density component $\nu_{\rm tot}$ is the
sum of the bulge and disc luminosity-density components $\nu_{b}$ and
$\nu_{d}$, which are the deprojections (Section~\ref{sec:deprojection}) of the bulge and disc
surface-brightness profiles derived in Section~\ref{sec:bulge-disk-comps}.

The relative \chisquare{} and marginalized likelihood plots for this
model are shown in the upper left part of
Figure~\ref{fig:n307-likelihood-all-models}. The best-fit SMBH mass ($7 \pm 1
\times 10^{7} \Msun $) is rather low -- about a factor of four smaller
than what the \msigmarel{} would predict (see
Section~\ref{sec:smbh-in-n307}) -- though by itself not obviously
implausible. The stellar $M/L$ is apparently quite well-defined.

\subsection{Model A+DM: SMBH + Single Stellar Component + DM Halo}
\label{sec:model-a-plus-dm}

Model~A+DM is Model~A with the addition of a dark-matter halo, so that the mass model is
\begin{equation}
\rho \; = \; \mbh \, \delta(r) \: + \: \mlt \, \nu_{\rm tot} \: + \: \rho_{\rm DM}.
\end{equation}
The DM halo is a standard spherical cored logarithmic model
\citep[e.g.,][]{binney-tremaine87}, with a density profile given by 
\begin{equation} 
\rho_{\rm DM}(r) \; = \; \frac{V_{h}^{2}}{4 \pi G} \frac{3 r_{h}^{2} + r^{2}}{(r_{h}^{2} + r^{2})^{2}}, 
\end{equation} 
where $r_{h}$ is the core radius (inside of which the density slope is
constant) and $V_{h}$ is the asymptotic circular velocity. Previous
studies modeling early-type galaxies with DM haloes have found that
similar results are obtained for both cored logarithmic and NFW DM halo
models \citep{thomas05,thomas07b}. Schwarzschild modeling of SMBH masses
including DM haloes have also found that the results do not depend
strongly on the specific DM halo model used
\citep{gebhardt09,mcconnell11a}.

The lower left part of Figure~\ref{fig:n307-likelihood-all-models} shows
relative \chisquare{} and marginalized likelihood plots for Model~A+DM.
The SMBH mass ($2.0 \pm 0.3 \times 10^{8} \Msun$) is about three times
larger than the Model~A value; the stellar $M/L$ ratio is about 15\%
smaller ($\mlkt = 1.1 \pm 0.1$).

Model~A+DM required almost 30 times the computational effort of Model~A.

\subsection{Model B: SMBH + Bulge + Disc}
\label{sec:model-b}

Model B is similar to Model~A except that there are \textit{two} stellar-density
components in the mass model, each with its own $M/L$ ratio, so the mass model is
\begin{equation}
\rho \; = \; \mbh \, \delta(r) \: + \: \mlb \, \nu_{\rm b} \: + \: \mld \, \nu_{\rm d},
\end{equation}
where $\nu_{\rm b}$ and $\nu_{\rm d}$ are the bulge and disc
luminosity-density models, respectively. These two components are
deprojections (Section~\ref{sec:deprojection}) of the bulge and disc
surface-brightness models derived in Section~\ref{sec:bulge-disk-comps}.

Relative \chisquare{} and marginalized likelihood values for this model
are shown in the upper right part of
Figure~\ref{fig:n307-likelihood-all-models}. The disc-component $M/L$
value (\mld) is implausibly high ($1.9 \pm 0.1$); however, the
\textit{bulge} $M/L$ value ($1.1 \pm 0.1$) is lower than the global
$M/L$ of Model~A, and is in fact identical to the global $M/L$ value of
Model~A+DM. The SMBH mass ($3.0 \pm 0.5 \times 10^{8} \Msun$) is about
50\% larger than that from Model~A+DM, but still consistent with the
latter at the $\sim 2-\sigma$ level; it is over four times larger than the
Model~A value.

Model~B required about six times the computational effort as Model~A,
but only one-fifth that of Model~A+DM.

\begin{table*}
\centering
\begin{minipage}{126mm}
\caption{NGC 307: Best-Fit Results from Dynamical Modeling}
\label{tab:n307-results}
\begin{tabular}{lccccccr}
\hline
Model Name & $M_{\rm BH}$     & \mlt          & \mlb{}        & \mld{}        & $r_{h}$ & $V_{h}$ & $t_{\rm comp}$ \\
           & ($10^{8} \Msun$) &               &               &               & (kpc)   & (\kms)  & (CPU h) \\
(1)        & (2)              & (3)           & (4)           & (5)           & (6)     & (7)     & (8) \\
\hline
A          & $0.70 \pm 0.1$   & $1.3 \pm 0.1$ & ---           & ---           & ---     & ---          &   1200$^{1}$ \\
A+DM       & $2.0 \pm 0.5$    & $1.1 \pm 0.1$ & ---           & ---           & $> 4.5$ & $200 \pm 20$ &   33000 \\
B          & $3.0 \pm 0.5$    & ---           & $1.1 \pm 0.1$ & $1.9 \pm 0.1$ & ---     & ---          &    7400 \\
B+DM       & $2.2 \pm 0.6$    & ---           & $1.1 \pm 0.1$ & $1.0 \pm 0.1$ & $> 5.6$ & $260 \pm 30$ &  200000 \\
\hline 
\end{tabular}

\medskip

Final results of dynamical modeling for NGC~307. (1) Model Name --
see Table~\ref{tab:n307-models}. (2) SMBH mass. (3) $K$-band stellar
$M/L$ ratio for combined stellar component. (4) $K$-band stellar $M/L$
ratio for bulge component. (5) $K$-band stellar $M/L$ ratio for disc
component. (6) DM halo radius. (7) DM halo circular velocity. (8) Total
computation time in CPU hours. Notes: 1. Some additional time was spend
exploring the low-\mbh{} part of parameter space for this model, since
the standard parameter-grid exploration yielded only an upper limit on
\mbh.

\end{minipage}
\end{table*}

\begin{table}
\caption{NGC 307: Comparison of Models}
\label{tab:n307-aic}
\begin{tabular}{lcccc}
\hline
Model &     Q1   &     Q2   &      Q3   &   Q4   \\
\multicolumn{5}{c}{AIC values} \\
A     &   728.6  &   574.9  &   1075.2  & 964.1  \\
A+DM  &   692.7  &   508.7  &   1050.9  & 918.3  \\
B     &   694.9  &   518.4  &   1051.4  & 917.5  \\
B+DM  &   694.3  &   504.6  &   1050.4  & 916.0  \\

\multicolumn{5}{c}{BIC values} \\
A     &   738.7  &   585.0  &   1085.5  & 974.5  \\
A+DM  &   717.9  &   534.0  &   1076.6  & 941.8  \\
B     &   710.1  &   533.6  &   1066.9  & 933.0  \\
B+DM  &   719.6  &   529.9  &   1076.1  & 941.8  \\
\hline 
\end{tabular}

\medskip

Comparison of different models fit to the NGC~307 data. Since each model
was fit to the kinematic data in each quadrant separately, we list the
corresponding Akaike Information Criterion (AIC) and Bayesian
Information Criterion (BIC) values for each quadrant (Q1, Q2, Q3, Q4)
separately. (See Figure~\ref{fig:sinfoni-kinematics} for how the
quadrants were specified.) Lower values of AIC or BIC indicate better
matches between model and data for a given quadrant (accounting for
differences in the number of free parameters).

\end{table}

\subsection{Model B+DM: SMBH + Bulge + Disc + DM Halo}
\label{sec:model-b-plus-dm}

Model~B+DM is the most complex model we consider. It is the same as Model~B
except that there is
also a DM halo, so the that the mass model is
\begin{equation}
\rho \; = \; \mbh \, \delta(r) \: + \: \mlb \nu_{\rm b} \: + \: \mld \nu_{\rm d} \: + \: \rho_{\rm DM}.
\end{equation}
The DM halo is the same spherical cored logarithmic model as we use in Model~A+DM.
The combined model
thus has a total of \textit{five} free parameters: \mbh, \mlb, \mld, $r_{h}$, and
$V_{h}$.

\begin{figure*}
\begin{center}
\includegraphics[scale=0.85]{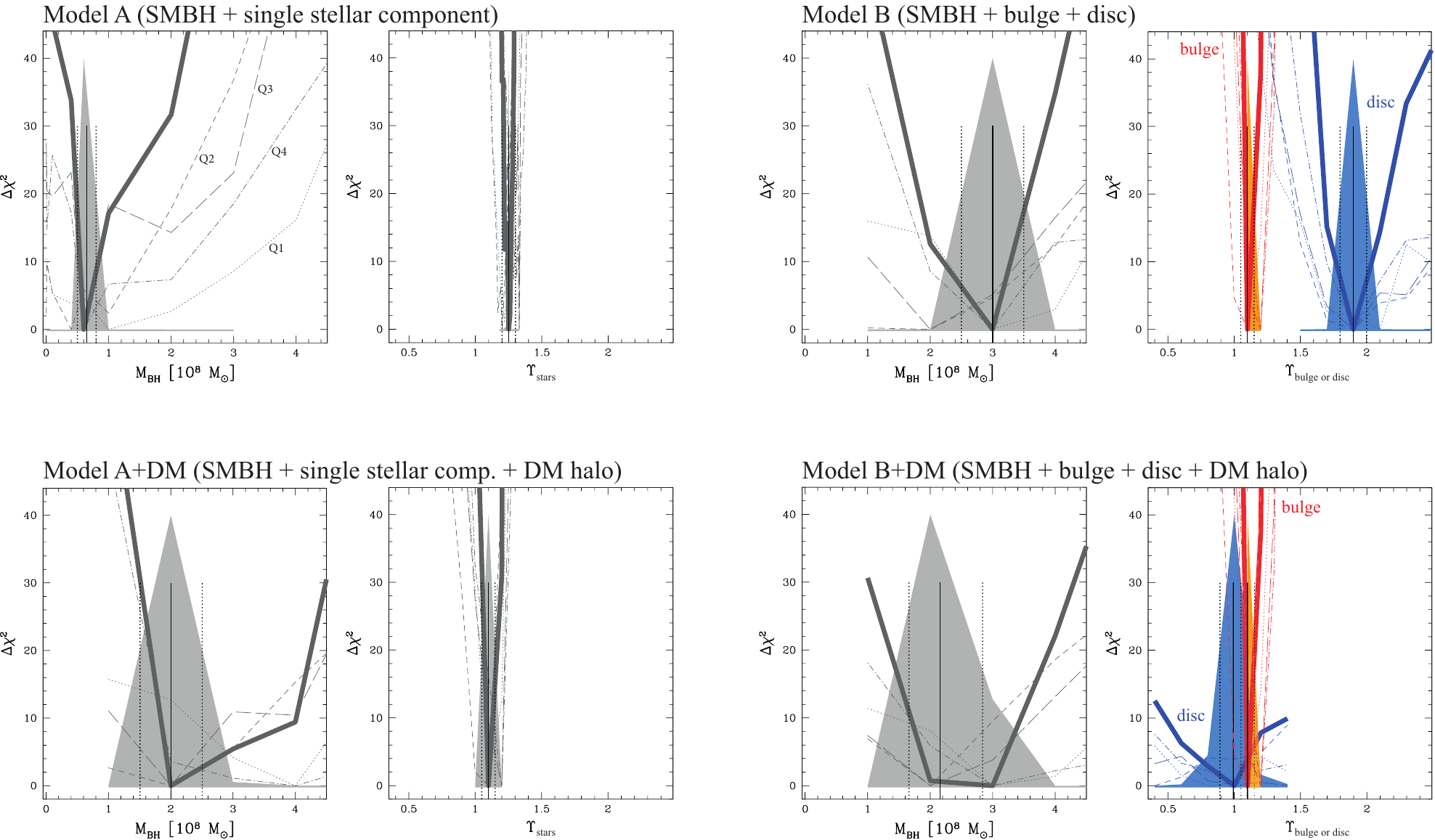}
\end{center}

\caption{ Relative \chisquare{} and marginalized likelihood plots for
the dynamical modeling of NGC~307, comparing all four general models.
For each model, we show $\Delta \chi^{2} = \chi^{2} - \chi^{2}_{\rm
min}$ values with gray, red, or blue lines, with thin lines showing
values for modeling of individual quadrants of kinematic data (Q1 =
dotted, Q2 = short-dashed, Q3 = long-dashed, Q4 = dot-dashed) and thick
lines showing the sum over all four quadrants. Likelihood values
(combining results for all four quadrants) are indicated by the gray,
blue, or orange shading; the likelihoods are scaled to an arbitrary
maximum value of 40 in each panel. Vertical solid lines mark best-fit
values for each parameter and vertical dashed lines indicate 68\%
confidence intervals. For each model, the left-hand panels show black
hole mass, while the right-hand panels show stellar $M/L$ values for the
single-stellar components of Models~A and A+DM, or for the bulge [red]
and disc [blue] components of Models~B and B+DM. Upper left pair of
panels: Model~A (SMBH + single stellar component). Lower left pair of
panels: Model~A+DM (same as Model~A, but with DM halo added). Upper
right pair of panels: Model~B (SMBH + separate bulge and disc stellar
components). Lower right pair of panels: Model~B+DM (same as Model~B,
but with DM halo added).\label{fig:n307-likelihood-all-models}}

\end{figure*}

The lower right part of Figure~\ref{fig:n307-likelihood-all-models}
shows relative \chisquare{} and marginalized likelihood values for the
SMBH mass and the $M/L$ ratios for the bulge and disc components.  The
best-fit \mbh{} value ($2.2 \pm 0.6 \times 10^{8} \Msun$) is in between
the best-fit values from Model~A+DM and Model~B, and is more than
three times larger than the best-fit value from Model~A. The bulge $M/L$
value is identical to the value in Model~B (and the global $M/L$ ratio
of Model~A+DM). The \textit{disc} $M/L$ value ($1.0 \pm 0.1$) is only
about 60\% of the value in Model~B, and is thus now \textit{lower} than
the bulge $M/L$ ratio, in qualitative agreement with our spectroscopic
analysis (Section~\ref{sec:stellar-pop}).

\begin{figure*}
\begin{center}
\includegraphics[scale=0.85]{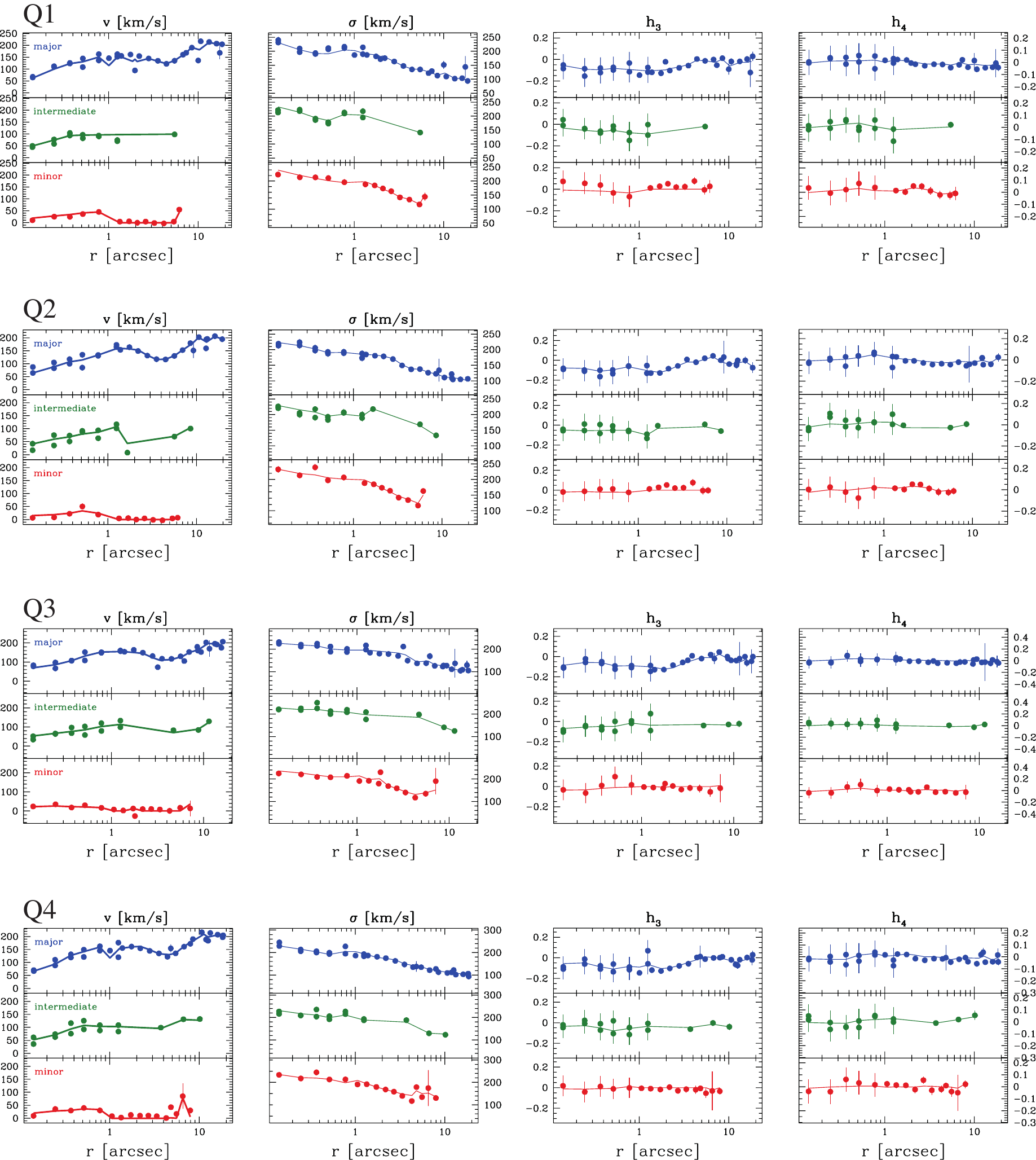}
\end{center}

\caption{Comparison of Gauss-Hermite kinematic data (points) and
best-fitting model predictions from Model~B+DM (lines) for kinematic
data from the four quadrants (Quadrants 1--4 are presented one per row,
from top to bottom). For each quadrant, we plot $V$, $\sigma$, $h_3$,
and $h_4$ extracted along the major (upper sub-panels, blue),
intermediate (middle sub-panels, green), and minor (lower sub-panels, red)
axes, along with corresponding values from the best-fit model for that
quadrant. Note that models were fit to the full LOSVDs from the 2D data
in each quadrant, \textit{not} to the Gauss-Hermite moments plotted in
the figure. \label{fig:n307-kinematics-models-vs-data}}

\end{figure*}

Since we consider this the best model for NGC~307 (see discussion
below), we show details of the fits to the kinematic data in
Figures~\ref{fig:n307-kinematics-models-vs-data}. This compares the
predicted stellar kinematics from the best-fit model with the kinematic
data from each of the four quadrants; note that for simplicity we show
Gauss-Hermite moments -- $V$, $\sigma$, $h_3$, and $h_4$ -- derived from
the full LOSVDs.

With a total of five free parameters, Model~B+DM required 200,000 CPU
hours of computational time -- six times that of the other DM-halo model
(Model~A+DM) and almost 30 times that of Model~B.

\begin{figure}
\begin{center}
\hspace*{-5mm}
\includegraphics[scale=1.15]{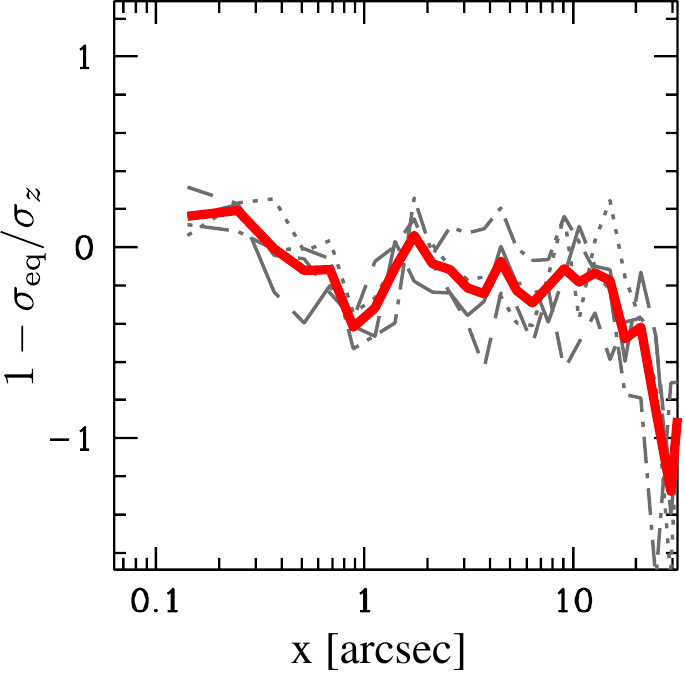}
\end{center}

\caption{Plots of orbital equatorial anisotropy for the orbits in
our preferred best-fit model (Model~B+DM). We show $1 -
\sigma_{eq}^{2}/\sigma_{z}^{2}$, where $\sigma_{eq}$ is the equatorial
dispersion ($\sigma_{eq}^2 = \sigma_{r}^2 + \sigma_{\phi}^2$) and
$\sigma_{z}$ is the vertical dispersion. The thin black lines represent
results from models to the individual quadrants; the thick red line is
the mean. The overall trend is for isotropic dispersion out to a radius
of $\sim 10\arcsec$, with increasingly strong equatorial anisotropy (as
expected for a rotationally dominated disc) outside.
\label{fig:anisotropy}}

\end{figure}

\subsection{Comparison and Summary of Modeling}

The effect of \textit{not} including a DM halo in the
single-stellar-component case (Model~A) is easily understood, because it
is similar to the effects seen for elliptical galaxies (always modeled
as single-stellar-component systems). Without a DM halo,
the stellar component needs a higher $M/L$ ratio ($\mlt = 1.3$) in order
to match the observed kinematics at large radii, where the (real) DM
halo starts to become significant compared to the stars. Since the
stellar $M/L$ ratio is the same at all radii, this effect also increases
the stellar mass in the inner regions of the galaxy, and so a lower SMBH
mass is needed to in order to match the observed kinematics there.

When a DM halo is added to Model~A (creating Model~A+DM), the effect is
fairly dramatic: although the stellar $M/L$ ratio decreases only
moderately (from 1.3 to 1.1), the SMBH is almost three times larger
($\mbh = 2.0 \pm 0.5 \times 10^{8} \Msun$). As is true for elliptical
galaxies modeled with DM haloes, the halo is able to replace the role of
the stellar component in accounting for the outer kinematics.
Consequently, the stellar component can acquire a lower value, and the
SMBH mass can correspondingly increase.

For Model B (the two-stellar-component model without DM halo), the disc
component is affected in a fashion similar to (but even stronger than)
that of the single stellar component in Model~A: in order to explain the
observed kinematics at large radii, the disc $M/L$ ratio is biased high
($\mld = 1.9$) to compensate for the absence of a DM halo. However, the
presence of a separate bulge stellar component -- which dominates the
stellar mass budget at small radii -- breaks the direct connection
between outer stellar $M/L$ ratio and SMBH mass that bedevils Model~A.
Instead, the bulge $M/L$ ratio and the SMBH mass can vary as needed to
better match the observed central kinematics. The result is a
\textit{lower} $M/L$ ratio for the bulge component and a higher mass for
the SMBH. The bulge $M/L$ value ($\mlb = 1.1 \pm 0.1$) is identical to
the global stellar $M/L$ value in Model~A+DM; the SMBH mass is only 50\%
higher. The only obvious problem with Model~B is the unrealistically
high $M/L$ ratio for the disc component -- almost twice the bulge $M/L$
ratio. This is directly contradicted by the stellar-population analysis
in Section~\ref{sec:stellar-pop}, which indicated that the disc $M/L$
ratio should be \textit{lower} than the bulge $M/L$ ratio.

Adding a DM halo to Model~B (Model~B+DM) primarily affects the disc
$M/L$ ratio: instead of there needing to be excess stellar mass at large
radii in order to explain the observed kinematics, mass can be shifted
into the DM halo component. The result is a much lower -- and much more
plausible -- $M/L$ ratio for the disc component of $\mld = 1.0$. Because
the outer stellar component remains decoupled from the inner component,
the effect on the bulge $M/L$ ratio and thus the SMBH mass is relatively
mild. In fact, the bulge $M/L$ ratio is unchanged from the Model~B value,
and the SMBH mass is in between the values for Model~A+DM and Model~B.

Because our kinematic data do not extend much beyond the
baryon-dominated inner regions of the galaxy, they cannot provide strong
constraints on the DM halo. In practice, fitting the two models with a
DM halo component (A+DM and B+DM) yields only lower limits on the halo
radius and somewhat discordant asymptotic velocities ($200 \pm 20$
\kms{} for A+DM, $260 \pm 20$ \kms{} for B+DM).

Do the models with extra components provide significantly better fits to
the data in a purely statistical sense? Since we are comparing multiple
models which are not simply nested (e.g., while Model~A is nested within
Model~A+DM, Models~B  and B+DM are not), direct comparisons of
\chisquare{} values is not valid. Instead, we look at more general
comparisons using information-theoretic statistics, which \textit{can}
be used to compare non-nested models that are fit to the same data.
Table~\ref{tab:n307-aic} compares the best fits of the different models
using the Akaike Information Criterion (AIC; see
Section~\ref{sec:2dfits}) and also the Bayesian Information Criterion
\citep[][]{schwarz78}. The AIC (actually the ``corrected'' AIC$_{\rm c}$
value) and BIC values are calculated using the \chisquare{} term from
Eqn.~\ref{eq:chi2}. As noted in Section~\ref{sec:2dfits}, lower values
of AIC (or BIC) indicate better fits; differences of $< 2$ are
insignificant, while differences of $> 6$ are considered strong evidence
that the model with the lower AIC or BIC is superior.

In this context, Model~A is clearly the worst model: its AIC values are
$\sim 9$--55 higher than those of the other models, and its BIC values
are 24--70 higher. The other models are practically indistinguishable
from each other in terms of AIC and BIC values. For example, only for
the Q2 value is Model~B+DM clearly superior to Model~B. The BIC values
actually favor Model~B over Model~B+DM ($\Delta$BIC $\approx 9$) for all
datasets except Q2.

What this shows is that our kinematic data are insufficient to
clearly discriminate between Models~A+DM, B, and B+DM. The data, for
example, do not allow us to distinguish between the case of a massive disc
with no DM halo (Model~B) and the case of a low-mass disc with a DM halo
(Model~B+DM).

\subsection{Variations: Testing the Sensitivity of Fits to Bulge/Disc Decompositions}

The method we use for generating the luminosity-density models
involves a 2D bulge-disc decomposition (with multiple sub-components for
the ``disc''). Uncertainties in this process translate into
uncertainties in the amount of light assigned to different components.
Since for Models~B and B+DM we assign potentially different $M/L$ ratios
to the bulge and disc components, the decomposition uncertainties could,
in principle, affect our derived $M/L$ ratios and SMBH masses.

To test how much variations in the bulge/disc decomposition might
actually affect the derived model parameters, we ran additional fits of
Model~B using divergent versions of our bulge/disc decompositions
corresponding to 1-$\sigma$ deviations from the best fit. This is
described in more detail in Appendix~\ref{app:decomp-variations}. The
results can be summarized as effectively \textit{no} discernable changes
in the black hole mass or stellar $M/L$ ratios for fits using $\pm
1$-$\sigma$ variations on the best-fit decomposition, so we conclude
that our results are not significantly affected by uncertainties in the
decomposition.

\subsection{Which Model Is Best? Accuracy Versus Efficiency and Strategies
for Modeling}

We are left with three models -- A+DM, B, and B+DM -- which are
approximately equally good at fitting the kinematic data. How can we
choose among them? From a general astrophysical perspective, Model B+DM
should be the most correct (or least wrong) model, since it allows for
both the possibility of different bulge and disc stellar $M/L$ ratios
(something we expect from both our general understanding of disc galaxy
evolution and from the spectroscopic evidence for NGC~307 itself)
\textit{and} the existence of a separate DM halo (something we expect
for all galaxies). The fact that the derived bulge and disc $M/L$ ratios
for Model~B+DM qualitatively agree with the spectroscopic results
(slightly higher in the bulge-dominated region, lower in the disc outside;
Section~\ref{sec:stellar-pop}) is further reason to prefer it over the
other models, although given the uncertainties in $M/L$ ratios, its
superiority relative to the Model~A+DM is not
statistically significant. Although Model~B allows for
different $M/L$ ratios in the bulge and disc regions, its agreement with
the spectroscopic analysis is actually worse, because it has a disc
$M/L$ ratio that is \textit{higher} than the bulge value. Moreover, its
disc $M/L$ value ($\mld = 1.9$) is too high to be physically plausible.

While the best model for NGC~307 is thus probably Model~B+DM, it
does have one practical drawback: the extensive computational time
required to evaluate it (200,000 CPU hours in our case).  The difficulty
posed by computational time for Schwarzschild modeling is illustrated
by the fact that recent studies which used the equivalent of our
Model~A+DM -- that is, including two DM halo parameters as part of the
fit, for a total of four free parameters -- have been devoted to one or
at most two galaxies only
\citep[e.g.,][]{gebhardt09,shen10a,jardel11,vandenbosch12,walsh15,
yildirim15,thomas16,walsh16}. Studies which included DM haloes for
more than two galaxies have avoided the expense of full parameter-space
searches by using fixed DM haloes in their models. \citet{schulze11}
specified fixed halo parameters based on galaxy luminosity, while
\citet{rusli13a} first fit three-parameter stars + DM halo models
(excluding the high-spatial-resolution data which probed the SMBH
region) to derive halo parameters as a function of \mlt, and then fit
SMBH + stars + DM halo models -- with only \mbh{} and \mlt{} as free
parameters -- to their full kinematic data.
Schwarzschild modeling with \textit{five} free parameters, as in our
Model~B+DM, has not previously been attempted, and is probably not (yet) a
practical approach for more than one or two galaxies at a time.

If we are interested in measuring reasonably accurate SMBH masses, and
potentially bulge $M/L$ ratios as well, for several galaxies at a time,
then Models~A+DM and B seem equally apropos: they yield SMBH masses
close to the Model~B+DM value and the same stellar $M/L$ ratio for the
bulge region as in Model~B+DM. Model~A+DM has a somewhat more accurate
SMBH mass, while Model~B is clearly the most \textit{efficient} way to
measure these quantities, since it has only three free parameters and
requires only $\sim 20$\% as much computational time as Model~A+DM.

\section{Discussion}

\subsection{The SMBH in NGC~307}\label{sec:smbh-in-n307}

Our preferred model (Model~B+DM) gives a SMBH mass of $\mbh = 2.2 \pm
0.6 \times 10^{8} \Msun$ for NGC~307. Given the previously published
central velocity dispersion of 205 \kms{} \citep{saglia16} and our
adopted distance of 52.8~Mpc, the diameter of the black hole's sphere of
influence would be $\approx 0.18\arcsec$. Our SINFONI observations had a
mean FWHM of $0.18\arcsec$, which means that our data (just)
resolve the SMBH's sphere of influence.

From the \msigmarel{} relation of
\citet{saglia16}\footnote{Specifically, the CorePowerEClassPC relation,
since NGC~307's status as an S0 with a classical bulge places it in that
particular sample.} we would derive an estimated SMBH mass of $2.67
\times 10^{8} \Msun$. Using the S{\'e}rsic model from our 2D
decomposition in Section~\ref{sec:2dfits}, the bulge of NGC~307 has
$M_{K} = -22.65$; with the bulge $M/L$ from Model~B+DM, this gives
$\mbulge = 2.97 \times 10^{10} \Msun$, so the SMBH is 0.74\% of the
bulge mass. The predicted SMBH mass from the CorePowerEClassPC
\mbulgerel{} relation in \citet{saglia16} would be $1.40 \times 10^{8}
\Msun$. The SMBH in NGC~307 is thus within $\sim 30$--40\% of what the
\msigmarel{} and \mbulgerel{} relations would
predict,\footnote{Differing from the predictions in $\log \mbh$ by only
0.08 and 0.20 dex, respectively, as compared with the measured RMS of
0.41 and 0.45 for the fits in \citet{saglia16}.} and is therefore quite
unexceptional.\footnote{Note that preliminary SMBH and bulge masses for
this galaxy ($\mbh = 4.0 \pm 0.05 \times 10^{7} \Msun$, $\mbulge = 3.2
\pm 0.4 \times 10^{10} \Msun$) were actually used to construct the
relations in \citet{saglia16}, but since NGC~307 was only one of 77
galaxies in the CorePowerEClassPC subsample, it did not have a strong
effect on the derivation of the relation.}

\subsection{Implications for SMBH Measurements in Disc Galaxies} 

Our analysis of NGC~307 suggests that attempts to measure SMBH masses in
disc galaxies via stellar-dynamical modeling can suffer from the same
limitations that have been found for elliptical galaxies. Specifically,
modeling a disc galaxy with just a single stellar
component (with a uniform $M/L$ ratio) and a SMBH can lead to 
underestimated SMBH masses and overestimated stellar $M/L$ ratios. This
can be alleviated by subdividing the stellar model into bulge and disc
components (increasing the number of free parameters to three), or by
adding a DM halo to the single-stellar-component model (increasing the
number of free parameters to four). The best approach is clearly to
model multiple stellar components \textit{and} a DM halo, but this is
computationally very expensive, since it involves five free parameters
rather than three or four.

Schwarzschild modeling of disc galaxies using a single stellar component
and no DM halo does not \textit{always} lead to biased SMBH mass
measurements, as the case of NGC~4258 shows. \citet{siopis09} obtained a
SMBH mass measurement for that galaxy which differed by only $\sim 15$\%
from the very high quality maser measurement. \citet{rusli13a} showed
that biases to SMBH measurements without DM haloes in elliptical galaxies
could be avoided if the inner kinematic data used in the modeling had
sufficiently high spatial resolution -- ideally several times better
than the SMBH's sphere of influence. Since the \textit{HST} STIS
kinematic observations used for the Siopis et al.\ analysis of NGC~4258
(FWHM $\approx 0.1\arcsec$) significantly over-resolved the SMBH sphere
of influence ($d \approx 0.7\arcsec$, assuming $\sigma = 115$~\kms, $D =
7.27$ Mpc, and $\mbh = 3.8 \times 10^{7} \Msun$ from the compilation in
\citealt{saglia16}), Schwarzschild modeling of the SMBH mass would
understandably be insensitive to the lack of a DM halo.

Based on our findings, and by analogy with the results for elliptical
galaxies, it seems plausible that disc galaxies where the SMBH sphere of
influence is only just resolved -- or is \textit{under}-resolved -- would be the
likeliest candidates to have biased SMBH measurements when modeled with
only one $M/L$ ratio and no DM halo. From the recent compilation of
\citet{saglia16}, there are eighteen disc galaxies with SMBH masses from
Schwarzschild modeling.\footnote{Or seventeen if NGC~524 is considered
to be an elliptical galaxy.}\footnote{Since the details of the
measurements for NGC~4736 and NGC~4826 -- listed in \citealt{kormendy13}
-- have not yet been published, we do not consider them.} Four of these
have been modeled with a single stellar component \textit{and} a DM halo
\citep{schulze11,walsh16}, and another four were modeled with two
stellar components \citep{davies06,nowak10,rusli11}. Of the remainder,
we can identify two for which the FWHM of the kinematic observations is
$\ga$ the diameter of the sphere of influence: NGC~1023
(\citealt{bower01}; FHWM = 0.2\arcsec, $\dsoi = 0.16\arcsec$) and
NGC~2549 (\citealt{krajnovic09}; FWHM = 0.17\arcsec, $\dsoi =
0.10\arcsec$). We suggest that those two galaxies in particular could
benefit from remodeling with multiple stellar components or with
DM haloes (or both).

\subsection{Stellar Orbital Structure}\label{sec:orbits}

Schwarzschild modeling produces a distribution of weights for the
different pre-calculated orbits in the model potential. From these, it is possible to learn something about the stellar orbital structure in the
best-fitting model. As we have done in past studies
\citep[e.g.,][]{thomas14,erwin15-composite}, we examine the radial trend
in orbital anisotropy. Specifically, we adopt the approach of Erwin et
al.\ and calculate an anisotropy parameter using the ratio of
planar/equatorial velocity dispersion $\sigma_{eq}$ to the vertical velocity
dispersion $\sigma_{z}$ (assuming cylindrical coordinates $R, \varphi, z$), where the mean dispersion in the equatorial plane is defined by
\begin{equation}
\sigma_{eq}^2 = (\sigma_{R}^2 + \sigma_{\varphi}^2)/2 \, .
\end{equation}
We compute the averages at each radius from orbits in angular bins that range from
$\theta = -23\degr$ to $\theta = +23\degr$ with respect to the equatorial plane.
The anisotropy $\beta_{\rm eq} = 1 - \sigma_{eq}^{2} / \sigma_{z}^{2}$ is $\sim 0$ for
isotropic velocity dispersion and $< 0$ for planar-biased anistropy; values
of $\sim -1$ are typical for the Galactic disc in the Solar neighborhood 
\citep[e.g.,][]{bond10}.

Figure~\ref{fig:anisotropy} shows that isotropy
($\beta_{\rm eq} \sim 0$) is the rule for $r \la 12\arcsec$. For $r \la
5\arcsec$, this is consistent with the evidence from the photometric
decomposition and the stellar-population analysis for a classical bulge.
The region $r \sim 5$--12\arcsec{} is outside the bulge, and so at first
glance it is puzzling that the velocity dispersion remains roughly
isotropic. However, $r \sim 12$ \textit{is} roughly where our
exponential-disc component begins to dominate the light (see
Figure~\ref{fig:n307-major-axis}). This suggests that the near-isotropy
between $\sim 5$ and 12\arcsec{} may be related to the weak bar or lens,
which contributes to the light in that radial range.
We note that although lenses are in general poorly studied, some
previous stellar-kinematic observations and models of barred
galaxies have suggested that lenses are kinematically hot, possibly dominated
by chaotic orbits or a large fraction of retrograde orbits
\citep[e.g.,][]{kormendy83,kormendy84a,pfenniger84b,teuben85,harsoula09}. Thus,
is it perhaps not surprising that the lens region in NGC~307 fails to show the
rotation-dominated anisotropy of a classical disc.

\section{Summary\label{sec:summary}}

We have presented 2D photometric decompositions, stellar kinematics from
adaptive-optics IFU and large-scale IFU and long-slit spectroscopy, and
dynamical modeling of the S0 galaxy NGC~307 with the aim of determining the mass
of its central SMBH.  We have paid particular attention to the effects
of modeling the stellar component as a single entity with one $M/L$
ratio versus modeling it as two sub-components (bulge and disc) with
independent $M/L$ ratios, and the effects of including a separate DM
halo in the modeling.

Our best estimate, from the model with a SMBH, separate bulge and disc
components, and a DM halo (Model~B+DM), is a black hole mass of $2.2 \pm
0.6 \times 10^{8} \Msun$, $K$-band bulge and disc $M/L = 1.1 \pm 0.1
\, \msunlsun$ and $1.0 \pm 0.1 \, \msunlsun$, respectively, and a DM halo
(spherical cored logarithmic model) with core radius $r_c > 5.6$~kpc
and circular velocity $V_{h} = 260 \pm 30$ \kms. The SMBH mass is within
$\sim$ 40\% of the predicted value from the  $\mbh$--$\sigma$ relation
(assuming $\sigma_{0} = 205$ \kms) and is $\approx 0.74$\% of the bulge
stellar mass, making NGC~307 entirely consistent with standard
SMBH-bulge relations. The $M/L$ ratios are qualitatively consistent with
single-stellar-population modeling of our long-slit spectroscopy, which
implies a higher $M/L$ in the bulge region.

Modeling the stellar kinematics with both stellar components but
\textit{without} the DM halo (Model~B) produces identical results for the
bulge $M/L$ ratio ($1.1 \pm 0.1 \, \msunlsun$) and a slightly higher SMBH
mass ($3.0 \pm 0.5 \times 10^{8} \Msun$). The \textit{disc} $M/L$ ratio
is significantly higher ($1.9 \pm 0.1 \, \msunlsun$), due to the fact that
the disc component has to be more massive to account for the effects of
the (missing) halo.  This approach requires only $\sim 4$\% of the
computational time as Model~B+DM.

Modeling with a \textit{single} stellar $M/L$ for both bulge and disc
plus a DM halo (Model~A+DM) yields a SMBH mass almost identical to that
of Model~B+DM ($2.0 \pm 0.5 \times 10^{8} \Msun$) and a combined stellar
$M/L = 1.1 \pm 0.1 \, \msunlsun$; the DM halo then has core radius $r_c
> 4.5$~kpc and circular velocity $V_{h} = 200 \pm 20$ \kms{}. The
computational time required for this model is $\sim 20$\% of the time
required for model B+DM, but about 4.5 times that for Model~B.

Finally, the simplest model, with a single stellar $M/L$ ratio and
\textit{no} DM halo, gives a much lower value for the SMBH mass ($7.0
\pm 0.1 \times 10^{7} \Msun$) and a higher stellar $M/L$ ratio ($1.3 \pm
0.1 \, \msunlsun$), because the necessity of accounting for the DM halo
drives the stellar $M/L$ ratio to high values, increasing the stellar
mass everywhere and reducing the amount of mass that can be assigned to
the SMBH. This model is also clearly worse than the others in terms of
how poorly it fits the kinematic data.

This suggests that dynamical modeling of disc galaxies can yield
reasonably accurate measurements of SMBH masses and bulge $M/L$ ratios
without needing the additional computational time of including a DM halo
-- \textit{if} a separate disc component with its own $M/L$ ratio is
included, though the disc $M/L$ ratio will then almost certainly be
overestimated. Models that treat the entire galaxy as having a single
stellar $M/L$ ratio (\textit{without} a DM halo) can potentially
underestimate the SMBH mass by significant amounts, especially if the
kinematic data used do not overresolve the SMBH sphere of influence, as
has been previously found for elliptical galaxies. We suggest that
previous SMBH measurements for the S0 galaxies NGC~1023 and NGC~2549
should be revisited, since they were modeled using single stellar
components and no DM haloes, using kinematic data which probably does
not fully resolve their SMBH spheres of influence.

Our morphological and spectroscopic analysis of NGC~307, including 2D
decompositions, suggests that the galaxy has four distinct stellar
components: a compact central bulge with a metal-rich stellar population
($\approx 33$\% of the light), a weak bar or lens ($\approx 6$\%), an
exponential disc ($\approx 36$\%), and a rounder, luminous stellar halo
with slightly boxy isophotes ($\approx 25$\%) which is misaligned by
about 5\degr{} with respect to the disc and bulge. (In our
two-stellar-component dynamical modeling, we treated the disc + bar/lens
+ stellar halo as one component.) Using our best-fit $K$-band $M/L$
values, the estimated stellar masses for these components are $3.6
\times 10^{10} \Msun$ (bulge), $4.3 \times 10^{9} \Msun$ (bar/lens),
$2.7 \times 10^{10} \Msun$ (disc), and $1.9 \times 10^{10} \Msun$
(stellar halo), with a total stellar mass of $8.6 \times 10^{10} \Msun$.
The stellar halo is best understood as a separate component rather than
being simply the outer part of the bulge; this is consistent with recent
2D decomposition analyses of the Sombrero Galaxy, which indicate a bulge
+ stellar halo + disc model is a better match to the galaxy than a
single bulge component plus the disc \citep{gadotti12}.

\section*{Acknowledgements}

We thank VLT support astronomer Chris Lidman and telescope operator
Christian Esparza for their assistance during the VLT-SINFONI
observations. We also thank Karl Gebhardt for comments on an earlier
draft, and the anonymous referee for several useful comments and suggestions.

P.E. was partly supported by DFG Priority Programme 1177 (``Witnesses of Cosmic
History:  Formation and evolution of black holes, galaxies and their
environment''); S.P.R. acknowledges support from the DFG Cluster of Excellence
Origin and Structure of the Universe.

Based on observations made with ESO Telescopes at the La Silla and Paranal
Observatories under programme IDs 082.B-0037, 082.A-0270, and 084.A-9002.

This research made use of the NASA/IPAC Extragalactic Database (NED)
which is operated by the Jet Propulsion Laboratory, California
Institute of Technology, under contract with the National Aeronautics
and Space Administration.  It also made use of the Lyon-Meudon
Extragalactic Database (LEDA; part of HyperLeda at
http://leda.univ-lyon1.fr/).


\bibliographystyle{mn2e}

%
%

\appendix

\section{Effects of Variation in Bulge/Disk Decomposition on Dynamical Modeling}
\label{app:decomp-variations}

Our preferred dynamical models (Models~B and B+DM) have separate
bulge and ``disc'' (i.e., disc + bar/lens + stellar halo) stellar
components, each with its own $M/L$ ratio. There is
the possibility that uncertainties in the bulge/disc decomposition
(Section~\ref{sec:2dfits}) -- e.g., how much stellar light is assigned
to the bulge component -- might lead to uncertainties in the two stellar
$M/L$ ratios, and thus potentially also to uncertainties in the SMBH
mass. To investigate the possible effects of variations in the 
bulge-disk decomposition, we focused on the fits to the VLT-FORS1 image
(Section~\ref{sec:2dfits}). Using the bootstrapping facility in
\textsc{Imfit} (see Section~5 of \citealt{erwin15-imfit}), we generated
1000 resampled versions of the FORS1 image and fit each with the same
B+b+D+H model as we used for the main decomposition
(Table~\ref{tab:n307-decomp}). We then computed the $B/T$ values for
each best-fitting model. The standard deviation of the 1000 $B/T$ values
was $\sigma_{B/T} = 0.0046$ or $\sim 1.4$\% of the original best-fit
model's $B/T$ of 0.3265.

We selected two of the bootstrap-resampled fits, with $B/T$ values
equal to the best-fit value $\pm \sigma_{B/T}$. We then generated bulge
and disc model surface-brightness profiles and deprojected these to form
bulge and disc luminosity-density components, as in
Section~\ref{sec:deprojection}. Finally, we ran our dynamical modeling
process using these new stellar components. For the underlying general
dynamical model we used Model~B, which has SMBH, bulge, and disc
components. (We chose this general model because it contains separate
bulge and disc $M/L$ ratios but requires considerably less time to run
than Model~B+DM.)

\begin{figure*}
\begin{center}
\includegraphics[scale=0.87]{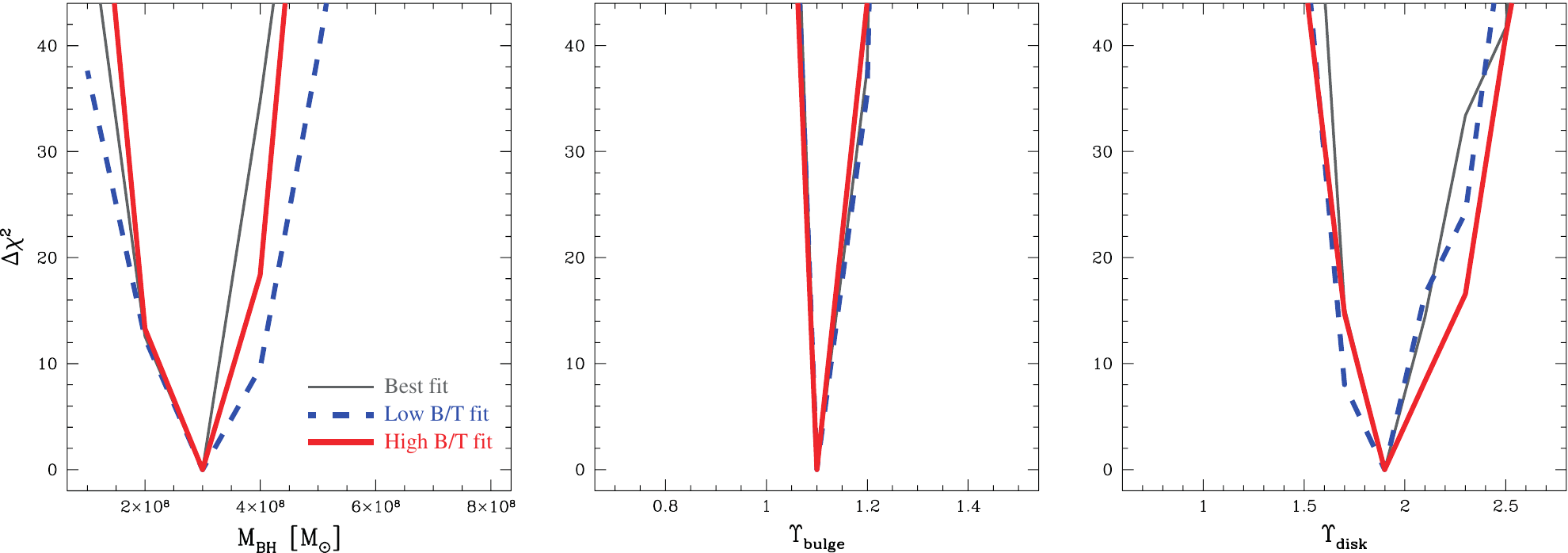}
\end{center}

\caption{Marginalized likelihood plots for Model~B (disc + bulge +
SMBH, no DM halo) using alternate bulge/disc decompositions, for the
SMBH mass (left), bulge $M/L$ (middle), and disc $M/L$ (right); see
Appendix~\ref{app:decomp-variations} for details. The dashed blue and
solid red curves are for the low- and high-$B/T$ decompositions,
respectively; the thinner dark grey curves are for the fit using the
best decomposition (same as in the upper-right panel of
Figure~\ref{fig:n307-likelihood-all-models}).
\label{fig:BtoT-variation-modeling}}

\end{figure*}

The results of the dynamical fits to these two decompositions are shown
in Figure~\ref{fig:BtoT-variation-modeling}. The SMBH mass is, within in
our admittedly somewhat coarse sampling, identical to our standard
best-fit results ($\mbh = 3.0 \pm 0.5 \, \Msun$) for Model~B (see
Table~\ref{tab:n307-results}). The bulge and disc $M/L$ ratios are also
identical ($\mlb = 1.1$, $\mld = 1.9$). We conclude that the nominal
uncertainties of our bulge-disc decomposition have minimal effect on the
results of our dynamical modeling, and in particular have negligible
effect on the SMBH mass determination.

\section{Long-Slit Stellar Kinematics for NGC~307}

The stellar kinematics (both major- and minor-axis) from our long-slit
spectra  of NGC~307 are presented in Table~\ref{tab:fors1-kinematics}.

\begin{table*}
\centering
\begin{minipage}{140mm}
\caption{VLT-FORS1 Stellar Kinematics}
\label{tab:fors1-kinematics}
\begin{tabular}{@{}lcccrrrrrrr}
\hline
PA    &     R     &   V     &   err  &  $\sigma$ &   err   &  $h_{3}$  &     err &  $h_{4}$  &    err  \\
(\degr)& (\arcsec) & (\kms) & (\kms) & (\kms)    & (\kms)  &           &         &           &         \\
(1)   &    (2)    &  (3)    &   (4)  &   (5)     &   (6)   &    (7)    &    (8)  &     (9)   &    (10)  \\
\hline
 78.1 &  $-$24.71 &  215.43 &  7.80  &  88.71    & 10.91   & $-$0.066  & 0.086   & $-$0.102  & 0.052 \\
 78.1 &  $-$22.68 &  196.03 &  4.40  &  96.95    &  5.86   & $-$0.054  & 0.053   & $-$0.075  & 0.040 \\
 78.1 &  $-$20.43 &  204.41 &  5.67  & 106.02    &  6.29   & $-$0.050  & 0.049   & $-$0.046  & 0.033 \\
 78.1 &  $-$18.56 &  204.15 &  4.33  &  90.08    &  5.05   &    0.044  & 0.043   & $-$0.064  & 0.030 \\
 78.1 &  $-$16.93 &  208.65 &  4.70  &  97.47    &  5.25   &    0.004  & 0.046   & $-$0.046  & 0.032 \\
\hline 
\end{tabular}

\medskip 

Binned kinematics for NGC~307 from our VLT-FORS1 observations. For the
major-axis spectrum (PA $= 78.1\degr$), negative radii are to the west;
for the minor-axis spectrum (PA $= 168.1\degr$), negative radii are to
the north. Note that these radii are the reverse of how the individual
profiles are plotted in Figures~\ref{fig:n307-fors1kin-major} and
\ref{fig:n307-fors1kin-minor}. Column 1: Position angle of slit (degrees
east of north). Column 2: Radius along slit (see above). Columns 3 and
4: Velocity and error (assuming systemic velocity = 3970 \kms). Columns
5 and 6: Velocity dispersion and error. Columns 7 and 8: Gauss-Hermite
$h_{3}$ coefficient and error. Columns 9 and 10: Gauss-Hermite $h_{4}$
coefficient and error. This is a preview of the full data table, which
is available online.

\end{minipage}
\end{table*}

\section{VIRUS-W IFU Kinematics for NGC~307}

The Voronoi-binned stellar kinematics from our VIRUS-W observations of
NGC~307 are presented in Table~\ref{tab:n307-virusw-kinematics}. The
definitions of the bins in terms of individual fibers, and the positions
of the latter on the sky, are presented in
Table~\ref{tab:n307-virusw-bins}.

\begin{table*}
\centering
\begin{minipage}{140mm}
\caption{VIRUS-W Stellar Kinematics}
\label{tab:n307-virusw-kinematics}
\begin{tabular}{@{}lrcccrrrrrr}
\hline
Bin   &    V    &   err    &  $\sigma$ &   err   &  $h_{3}$  &     err &  $h_{4}$  &    err  \\
      &( \kms) & (\kms)   & (\kms)    & (\kms)  &           &         &           &         \\
(1)   &    (2)  &  (3)     &   (4)     &   (5)   &   (6)     &    (7)  &    (8)    &     (9) \\
\hline
 0 & $-78.57$ & 2.38 & 209.57 & 2.73 &    0.037 & 0.008 & $-$0.006 & 0.009 \\
 1 &  $-7.64$ & 4.23 & 215.61 & 4.91 &    0.005 & 0.015 & $-$0.005 & 0.013 \\
 2 &   26.42  & 6.64 & 228.77 & 7.07 & $-$0.029 & 0.021 & $-$0.026 & 0.020 \\
 3 & $-69.16$ & 6.10 & 166.39 & 6.95 & $-$0.009 & 0.030 & $-$0.029 & 0.028 \\
 4 &   94.27  & 2.76 & 169.38 & 3.41 & $-$0.023 & 0.013 & $-$0.005 & 0.015 \\
\hline 
\end{tabular}

\medskip 

Binned stellar kinematics for NGC~307 from our VIRUS-W observations.
Column 1: Voronoi bin number (see Figure~\ref{fig:n307-virus-kinematics}
and Table~\ref{tab:n307-virusw-bins}). Columns 2 and 3: Velocity and
error (assuming systemic velocity = 3992 \kms). Columns 4 and 5:
Velocity dispersion and error. Columns 6 and 7: Gauss-Hermite $h_{3}$
coefficient and error. Columns 8 and 9: Gauss-Hermite $h_{4}$
coefficient and error. This is a preview of the full data table, which
is available online.

\end{minipage}
\end{table*}

\begin{table}
\caption{NGC 307: VIRUS-W Bin Assignments}
\label{tab:n307-virusw-bins}
\begin{tabular}{@{}lccc}
\hline
Fiber          & RA  & Dec &  Bin \\
(1)            & (2) & (3) &  (4) \\
\hline
 0       & 14.136430 & $-1.771680$  &  0 \\
 1       & 14.135780 & $-1.771290$  &  1 \\
 2       & 14.135780 & $-1.772130$  &  2 \\
 3       & 14.136480 & $-1.770000$  &  3 \\
 4       & 14.136470 & $-1.770840$  &  3 \\
\hline 
\end{tabular}
\medskip

Positions on the sky and Voronoi bins assignments for the VIRUS-W kinematics in
Table~\ref{tab:n307-virusw-kinematics}. Column 1: Fiber number. Column
2: Right Ascension (J2000) of fiber centre in decimal degrees. Column 3:
Declination (J2000) of fiber centre in decimal degrees. Column 4:
Voronoi bin that fiber was assigned to (see map in lower-left panel of
Figure~\ref{fig:n307-virus-kinematics}). This is a preview of the
full data table, which is available online.

\end{table}

\end{document}